\pdfoutput=1

\documentclass[review, a4paper,fleqn]{cas-dc}

\usepackage[numbers]{natbib}

\usepackage{amsmath}
\usepackage[ruled,linesnumbered]{algorithm2e}
\usepackage{pseudocode}
\usepackage{array}
\usepackage{subcaption}
\usepackage{stfloats}
\usepackage{graphicx}

\usepackage[nolist,nohyperlinks]{acronym}
\acrodef{BFS}[BFS]{\emph{Breadth-First Search}}
\acrodef{BIM}[BIM]{\emph{Building Information Modeling}}
\acrodef{AG}[SHAPE]{\emph{Spatial Human Accessibility graph for Planning and Environment analysis}}

\newcommand{\outedges}{\ensuremath{E^{\rightarrow}_D(v_i)}\xspace} 
\newcommand{\edgeSubgraph}{\ensuremath{D[\outedges]}\xspace} 
\newcommand{\wm}[2][e_{ij}]{\ensuremath{\Psi_{#2}(#1)}\xspace} 

\def\tsc#1{\csdef{#1}{\textsc{\lowercase{#1}}\xspace}}
\tsc{WGM}
\tsc{QE}
\tsc{EP}
\tsc{PMS}
\tsc{BEC}
\tsc{DE}

\newdefinition{rmk}{Def}

\begin{document}
\let\WriteBookmarks\relax
\def\floatpagepagefraction{1}
\def\textpagefraction{.001}
\shorttitle{Spatial Accessibility Graph}
\shortauthors{M Schwartz}

\title [mode = title]{Human Centric Accessibility Graph For Environment Analysis}                               

\author[1]{Mathew Schwartz}[type=editor,
                        auid=000,bioid=1,
                        orcid=0000-0003-3662-7203]
\cormark[1]
\ead{cadop@njit.edu}
\ead[url]{www.cadop.info, cadop@umich.edu}

\credit{Conceptualization of this study, Methodology, Software}

\address[1]{New Jersey Institute of Technology, Newark, NJ, USA}


\begin{abstract}
Understanding design decisions in relation to the future occupants of a building is a crucial part of good design. However, limitations in tools and expertise hinder meaningful human-centric decisions during the design process. In this paper, a novel Spatial Human Accessibility graph for Planning and Environment Analysis (SHAPE) is introduced that brings together the technical challenges of discrete representations of digital models, with human-based metrics for evaluating the environment. SHAPE: does not need labeled geometry as input, works with multi-level buildings, captures surface variations (e.g., slopes in a terrain), and can be used with existing graph theory (e.g., gravity, centrality) techniques. SHAPE uses ray-casting to perform a search, generating a dense graph of all accessible locations within the environment and storing the type of travel required in a graph (e.g., up a slope, down a step). The ability to simultaneously evaluate and plan paths from multiple human factors is shown to work on digital models across room, building, and topography scales. The results enable designers and planners to evaluate options of the built environment in new ways, and at higher fidelity, that will lead to more human-friendly and accessible environments.

\end{abstract}


\maketitle

\section{Introduction}\label{sec:intro}

Understanding design decisions in relation to the future occupants of a building is a crucial part of good design. However, technical limitations in tools and the need for an extensive understanding of human physiology prevent the ubiquity of meaningful human-centric decisions during the design process. While the topic of human movement has been discussed in the literature at length, there is no method that approaches the \textit{construction} of the graph-representation of a digital model with a physiological human factors perspective. 

When considering an environment designed for people, accessibility is one of the most fundamental criteria for evaluating the space's performance. While recent work in computer science and architecture have shown an increase in the use of data and quantified metrics for optimization or evaluation, there has been only a few works that truly consider human factors of an occupant at the physiological level. Although building design usually incorporates human factors, the types of calculations used are often limited in scope. For example, the term \textit{human comfort} is often used in architecture and urban design as a reference to the thermal, visual, and acoustic relationships of the environment and occupant~\cite{frontczak2011literature}. To improve analysis, and subsequently, the performance of a building for its occupants, we must be able to streamline the evaluation of a design, taking into account the complexities of human experience in aggregate.

Accessibility is not a binary value, and the difficulty associated with moving from one point to another must be included when considering the accessibility of a space. While \textit{walkability} as a term has various definitions~\cite{moura2017measuring,forsyth2015walkable}, it is often considered to be an aggregate of transportation, policies, mobility, and locations of points of interest~\cite{krambeck2006global,frank2010development}. This paper uses the term walkability as the ability to walk on/over, reflective of the~\textit{Environments Conditions}$\rightarrow$\textit{Traversable} theme illustrated in~\cite{forsyth2015walkable}. The term accessibility is used in this paper as a person's ability to access a space--from a human factors perspective. The various conditions surrounding this ability and the associated costs it has on a person, are vital considerations for human comfort. This is not to say that existing techniques of spatial analysis are not valid. Rather, this paper argues for additional human-centric data to be used, and proposes a method for doing so.

\begin{figure}
  \centering
    \includegraphics[page = 1, width=.45\textwidth]{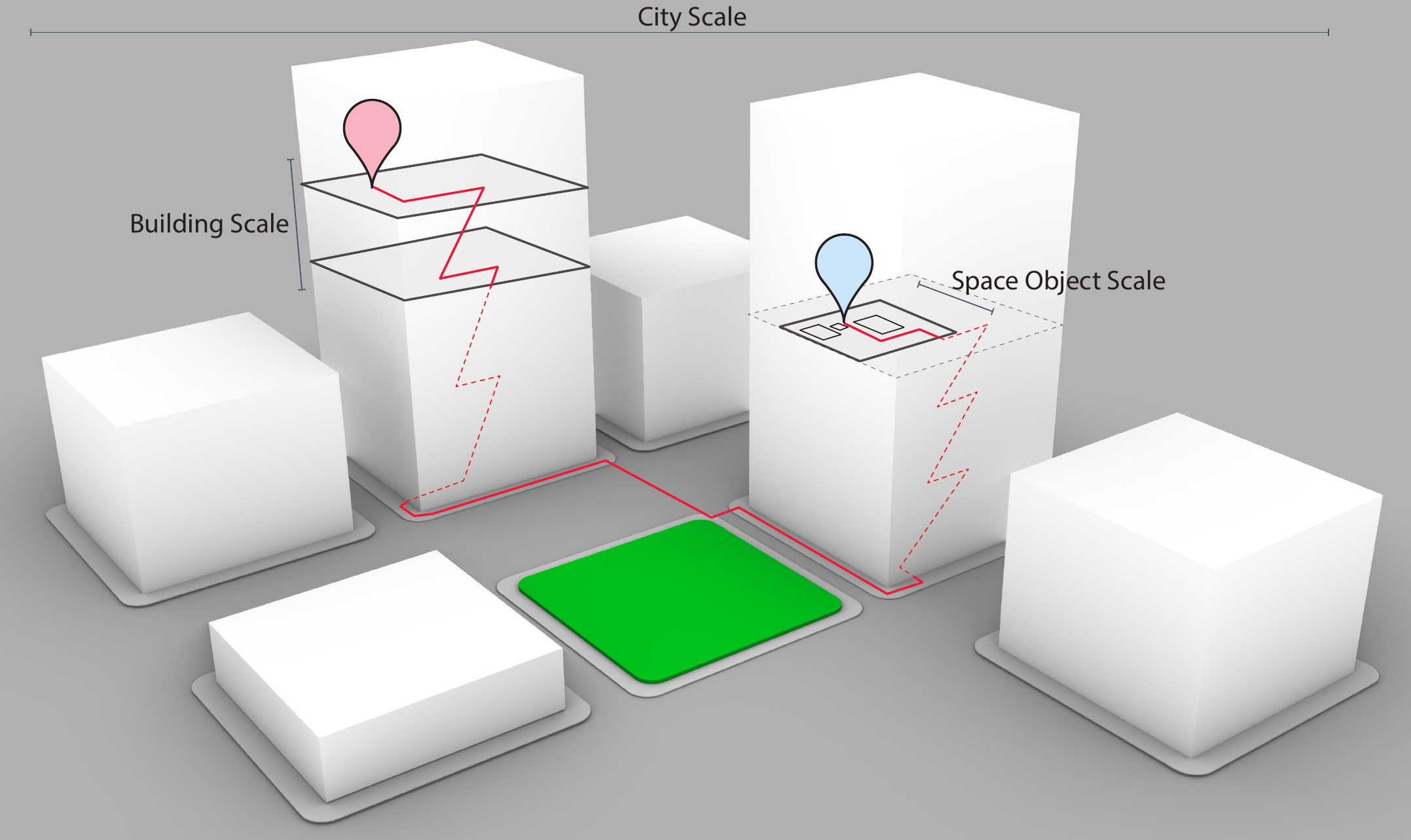}
  \caption{The three environmental scales of City, Building, and Object, are all integrated. Path analysis and accessibility is from one point to another, and includes numerous environmental conditions. (Adapted from~\cite{shin2019indoor})}
  \label{fig:envScales}
\end{figure}

While there are past works that evaluate the built environment with regard to human-based mobility such as slope and pavement quality~\cite{moura2017measuring}, the apparatus of the evaluation is often survey-based, checking if a space meets a given criteria that has been generalized or predefined. An important component of analyzing slope in an environment that is often lacking in survey data is the question:  "\textit{why is this slope bad?}". This disconnect can happen when the metric being used to analyze an environment relies on the comparison to building code and existing guidelines that create binary and abstracted measures (e.g., ~\cite{gamache2019mapping,bennett2009wheelchair}). Additionally, while automated code checking based on existing guidelines is valuable for assessing workflows and minimizing errors and time~\cite{eastman2009automatic}, the power of computation allows for a more complex relationship between accessibility and regulations. 

\begin{figure*}
  \centering
    \includegraphics[page=22, width=\textwidth]{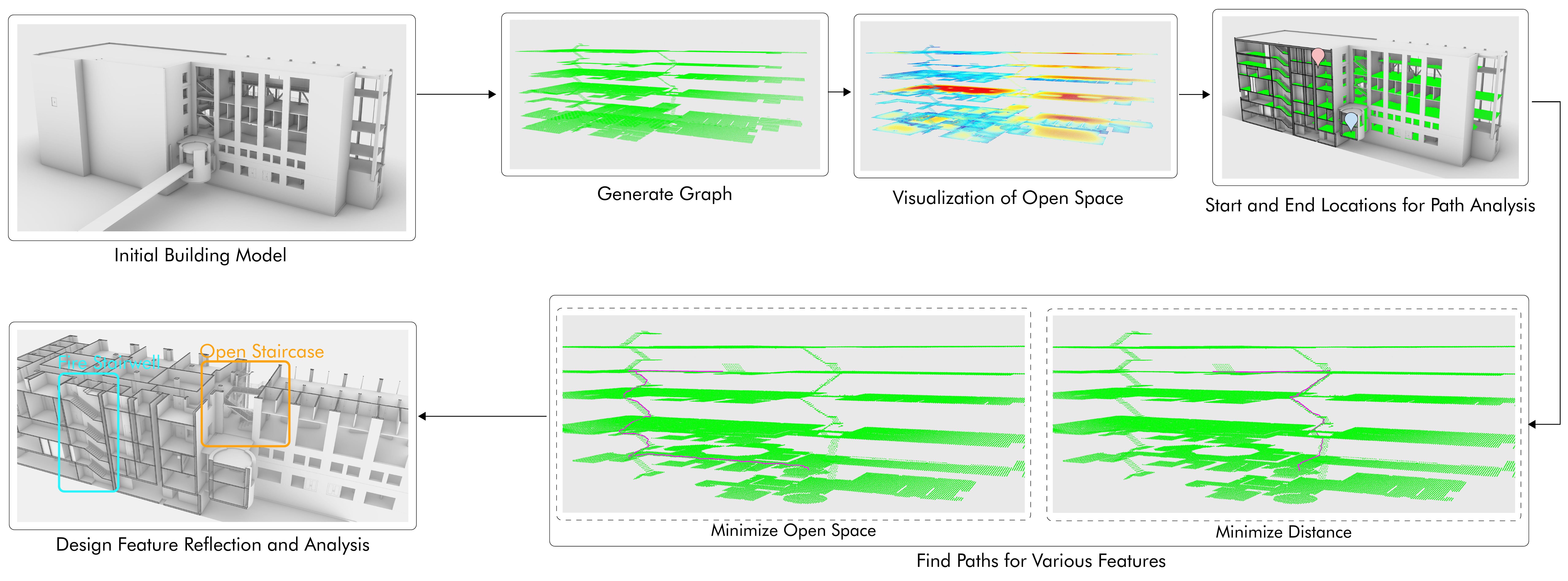}
  \caption{A multi-level building example of using the AG. Steps shown are 1) A starting model is given in the top left 2) the graph is generated finding accessible locations 3) a viewshed-type analysis is performed to find locations with maximal space 4) path planning start and end locations determined 5) generate paths with various criteria 6) reflection on paths generated by finding use cases for a given fire stairwell vs. the open staircase. }
  \label{fig:Overview_multiLevelPath}
\end{figure*}

Unless ultra-strict laws and regulations are enacted for explicitly limiting what is allowed to be an accessible path, it is not feasible for a designer to consider these conditions. Furthermore, this extreme restriction-based approach limits the design as well as opportunities to improve accessibility through innovation. Computation of human factors combined with automatic code checking provides an ideal platform for--performative rather than prescriptive--regulation. In order to achieve this goal, we must first solve the challenge of finding accessible locations, and subsequently, incorporating more than distances between the locations to properly represent the occupants experience.

In prior work, human-based accessibility analysis has been performed after the generation of a navigable surface. For example, a navigation mesh is first defined; then, a dense graph is generated within it, and a calculation is performed on accessibility factors thereafter; and finally, the graph is culled to the meaningful components. In this paper, a novel approach is presented for the first step -- defining the graph -- with a focus on human accessibility, enabling a scalable and information-rich dataset. The graph is constructed by rules dictated from accessibility related factors directly from the physiology and biomechanics literature. The algorithm is scalable from the object to urban scale~(Fig.~\ref{fig:envScales}) and can be used in multiple other analyses (from light intensity to fall probabilities) to inform the cost of accessibility. The approach allows for early design-stage feedback that can not only improve designs but also allows for quick iterations and reduced costs (Fig.~\ref{fig:Overview_multiLevelPath}). 

The two core contributions of the paper to the field are summarized as follows:
\begin{enumerate}
    \item A novel ray-casting based method for creating a grid-based weighted directed graph that enables early-stage feedback in the design process with respect to human experience and accessibility.
    \item An in-depth review of the current literature that categorizes and illuminates factors that should be included in path and circulation analysis both indoor and outdoor for ensuring human-centric built environments.
\end{enumerate}

The paper is organized as follows; Sec.~\ref{sec:humanFactors} presents a review of literature on human factors relating to accessibility and motivation for such an approach, Sec~\ref{sec:RelatedWork} provides an overview of the types of data structures and algorithms for evaluating 3D models, Sec.~\ref{sec:methodology} introduces a novel algorithm for parsing a building into a graph, Sec.~\ref{sec:implementation} demonstrates the presented work in a design context with various scales and environment types, and Sec.~\ref{sec:discussion} discusses limitations and future work.

\section{Human Factors}\label{sec:humanFactors}
Evaluating the accessibility of a space concerns not only Euclidean distance but also factors such as energy expenditure, speed, slips, trips, and falls. Likewise, analyzing an environment for human factors and comfort must include the tasks and movements to be performed in any given location (e.g., cutting vegetables or opening a cabinet). In this section, the basis for a weighted digraph built from an accessibility standpoint is introduced. 

While some laws and regulations exist around the type of floor surface of a building or urban environment such as the American Disabilities Act and its accessibility guidelines~\cite{board2002ada} for built environments--algorithmic methods for quantifying the accessibility of these environments are lacking. However, a large body of research exists that quantifies human biomechanics in relation to the built environment. By developing a method for interpreting a 3D model based on the way in which a human moves within a space, the literature based on human factors can be applied and used as a more informative metric than distance alone. 

From a biomechanical perspective, mobility within the built environment is impacted by surface types and geometry against physiological factors (e.g., age, weight, and disability). In Table~\ref{table:surf_type}, six types of surfaces are identified with a description and relevant reference to the biomechanics literature. These surface types apply to both the outdoor and indoor environments.    

\begin{table*}[width=1.9\linewidth,cols=3,pos=h]
\caption{Surface types and their significance in human mobility, with corresponding references. Surface material (e.g., sand or concrete) has an impact in all the examples; however the references focus on angles and variation.}\label{table:surf_type}
\begin{tabular*}{\tblwidth}{p{0.15\linewidth}p{0.6\linewidth}p{0.15\linewidth} }
\toprule
Surface Type & Description and Example & Reference \\
\midrule
Flat & Leveled and flat floor with little to no variation across the surface. & ~\cite{soule1972terrain}  \\
Ramp & Slope is along the progression axis. A person walks in the same direction as the incline. The balance is front to back. & \cite{rose1994energetics,minetti2002energy}  \\
Cross-slope & Slope is perpendicular to the progression axis (Fig.~\ref{fig:crossSlope}). A person walks perpendicular to the incline, forcing the body to adjust side to side. & \cite{dixon2010gait,cooper2011manual,pierret2014cardio}  \\
Staggered & Stairs and other flat surfaces that are offset and have a change in level, such as bricks that are protruding directly up in various heights (but not at an angle) or the offset of a curb to the sidewalk.  & \cite{novak2010stair,novak2016age,voloshina2013biomechanics,jacobs2016review}  \\
Uneven & Variations in curvature of the ground surface (Fig.~\ref{fig:unevenPhoto}), such as in most natural topography, especially mountains, and some poured concrete curved surfaces in modern design. & ~\cite{marigold2008age,schulz2011minimum}  \\
Uneven staggered & Faceted surfaces that are both angled and protruding, such as bricks in an old walkway with shifted ground underneath.  & \cite{dixon2018late}  \\
\bottomrule
\end{tabular*}
\end{table*}

\begin{figure}
  \centering
    \includegraphics[page = 7, width=.2\textwidth]{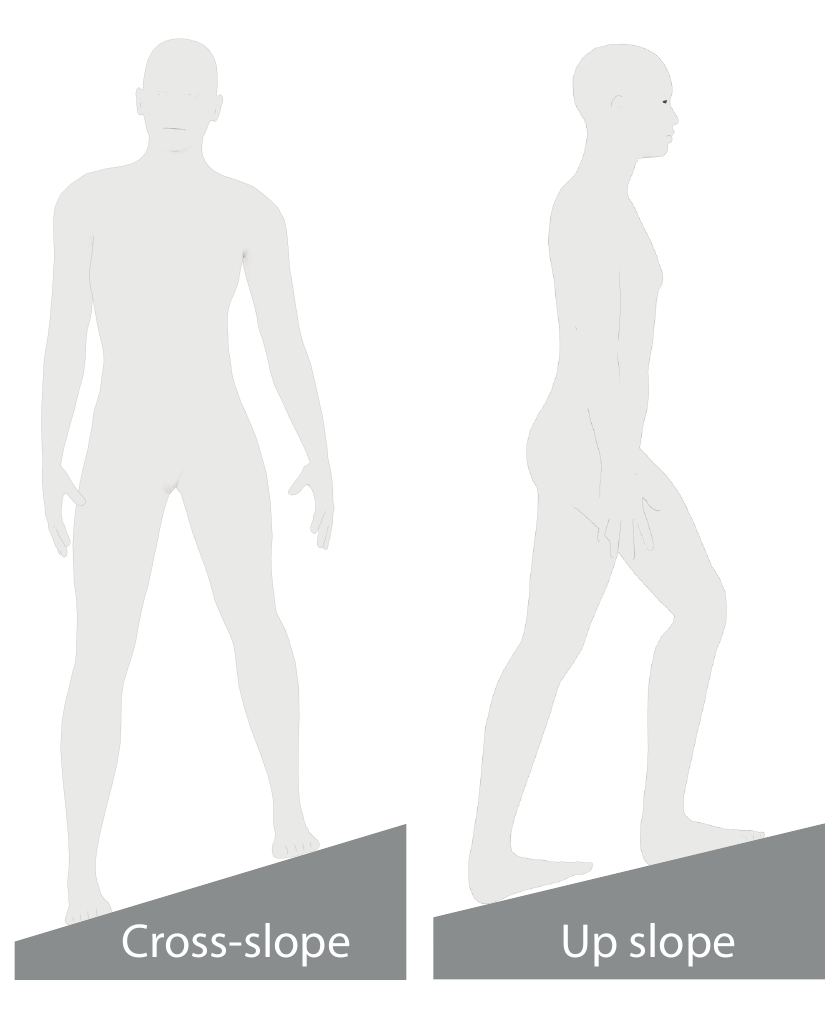}
  \caption{Cross-slope (left) is when the direction of movement is perpendicular to the direction of the slope. Cross-slope forces one side to have a shorter stance than the other when compared to a ramp (right) in which both sides equally vary during movement.}
  \label{fig:crossSlope}
\end{figure}

At the urban scale, the idea of a single brick out of place can seem irrelevant. However, these small details and aspects of the environment can significantly affect occupants.  Variation in brick height of a walkway can have a direct and meaningful impact on the ability for people to comfortably and safely move~\cite{dixon2018late}. Irregular surfaces (e.g., gravel) require significant adaptations in gait to reduce tripping, at the cost of speed, step length, and cadence~\cite{merryweather2011gait}. Furthermore, wheelchair propulsion energy costs increase significantly with surface type~\cite{koontz2005kinetic}. The net locomotive energy cost for a wheelchair on carpet compared to a tiled surface can increase by more than 70\%~\cite{glaser1981energy}. Surface resistance changes not only the energy required but also the speed and rate of movement (such as a nearly 50\% reduction in average velocity from tile to carpet)~\cite{cowan2009impact}. In~\cite{wolfe1977influence}, the authors explicitly mentioned the importance of floor type when designing a hospital owing to their findings in energy expenditure across concrete and carpeted floors. In~\cite{gonzalez1994energy}, the authors have made a similar point about the importance for designers to know this impact when designing homes. 

\begin{figure}[pos=h]
  \centering
  \includegraphics[width=2.5in]{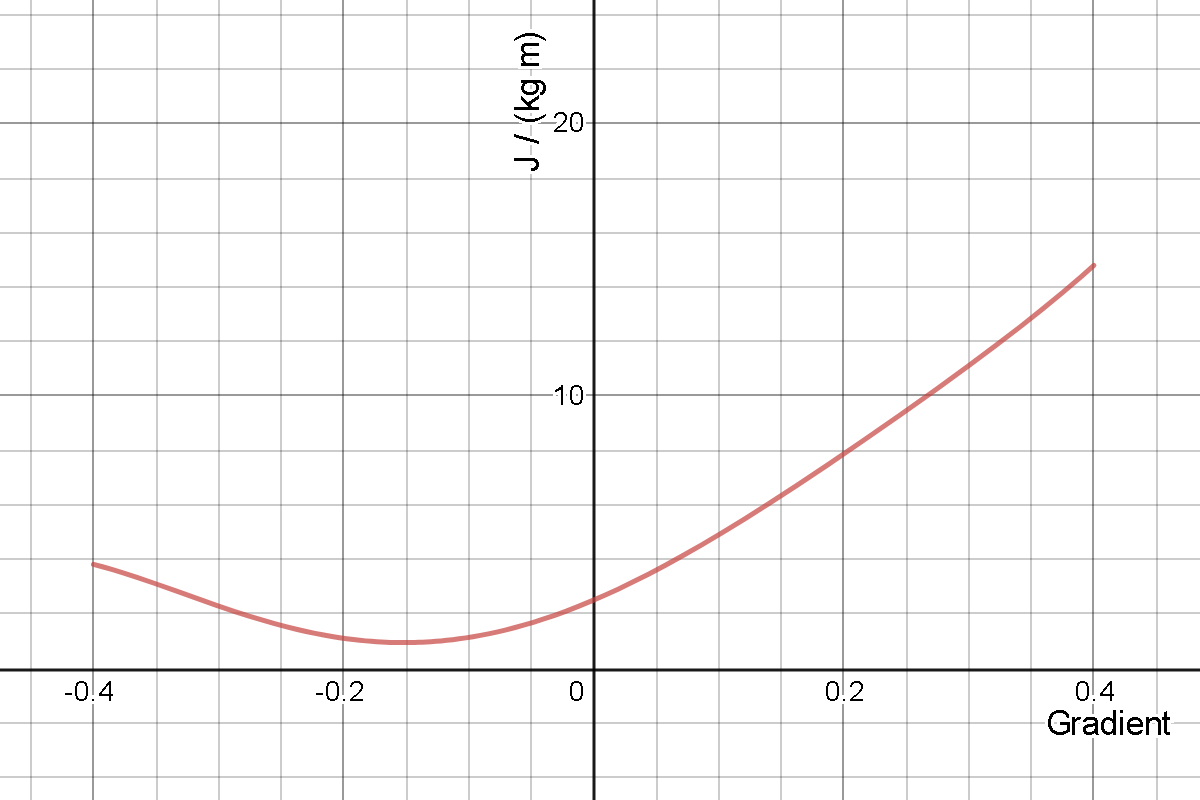}
  \caption{Energy cost of walking at various gradients as defined by the polynomial function in~\cite{minetti2002energy} and shown in Def~\ref{def:energyCost}. }
  \label{fig:energyGradient}
\end{figure}

When discussing disabilities and mobility an occupant using a wheelchair is often at the forefront of the discussion. It is important to remember the wide range of human mobility and assistive devices. For instance, canes and crutches may cause a reduction in mobility speed and energy up to 60\%~\cite{gonzalez1994energy}. The inclusion of this data for the analysis of travel times to amenities in the urban environment (walkability) or within a building would drastically change the results. 

The impact of a ramp on accessibility is not a simple relationship that every designer can instinctively work with. The increase in slope angle of a ramp is not linearly correlated to the energy increase it takes for a human to move up it~\cite{rose1994energetics,minetti2002energy}. As seen in Figure~\ref{fig:energyGradient}, the energy expenditure associated with the slope decreases in a slightly negative gradient. As the gradient negatively increases, the energy expenditure increases as well. Walking on a cross-slope (Fig.~\ref{fig:crossSlope}) of 6\textdegree ~has shown a threefold increase in mediolateral ground reaction force, which is an artifact of adaptations in movement to reduce falling~\cite{dixon2010gait}. The impact of a cross-slope for wheelchair propulsion is profound as well, with a 2\% cross slope increasing energy by over 30\%~\cite{brubaker1986effects}. ~\cite{voloshina2013biomechanics} studied a staggered block surface similar to bricks with a vertical variation up to 2.5 cm and observed a 28\% increase in energy expenditure.

In a study of adults under 60 years old, it was found that traumatic brain injuries from falls occurred most often in ground level falls (36\%), with the second most common being falls down stairs\footnote{For an extensive review of falling and stair negotiation, see~\cite{jacobs2016review}.} at 27\% ~\cite{friedland2014falls}. Stairs, especially in the urban environment, can significantly range in height, depth, and inclination. Accounting for not only the existence of a step or slope, but also the slope of steps, is a critical part of analyzing an occupant's path. Energy requirements have been shown to increase by 67\% from a 24\textdegree to a 42\textdegree incline. Moreover, the gait pattern is notably not a linear transition between walking and stair climbing, and it likely switches between the two at a certain threshold~\cite{riener2002stair}. Crutch ambulation is another example of the complexities in accessibility analysis. Studies have shown occupants using crutches on stairs had a twofold increase in energy expenditure, while stair usage was more efficient than ramps considering energy cost per vertical rise~\cite{gonzalez1994energy}. The caveat to this efficiency is when the ramp is less than 5\% it becomes the more efficient means of travel. 

While regulations exist for many stair dimensions, there is a need for a more cohesive analysis measurement for human mobility that takes into account a combination of environmental and spatial factors and conditions. This is evidenced by~\cite{hamel2005foot}, which demonstrated a decrease in ambient lighting from 300 lx to 3 lx (limit of civil twilight) led to young adults increasing foot clearance while almost all older adults did not. These small deviations (or lack there of) in human movement can lead to serious consequences (i.e., falling). 

It is important to note that the discussion around energy expenditure should not be limited to the simple issue of calories burned. In each case identified (e.g., stair negotiation or cross-slope wheelchair propulsion), an increase in energy expenditure can mean increased loads on joints and muscles, thus leading to injuries. When considering surface variation and cross-slope movements, a simplistic model of a line connecting spaces within the built environment is not sufficient. A graph-based approach in which the directed edges can be weighed based on a perpendicular edge (i.e., the cross slope) must be used.

\begin{figure}
  \centering
  \includegraphics[width=3in]{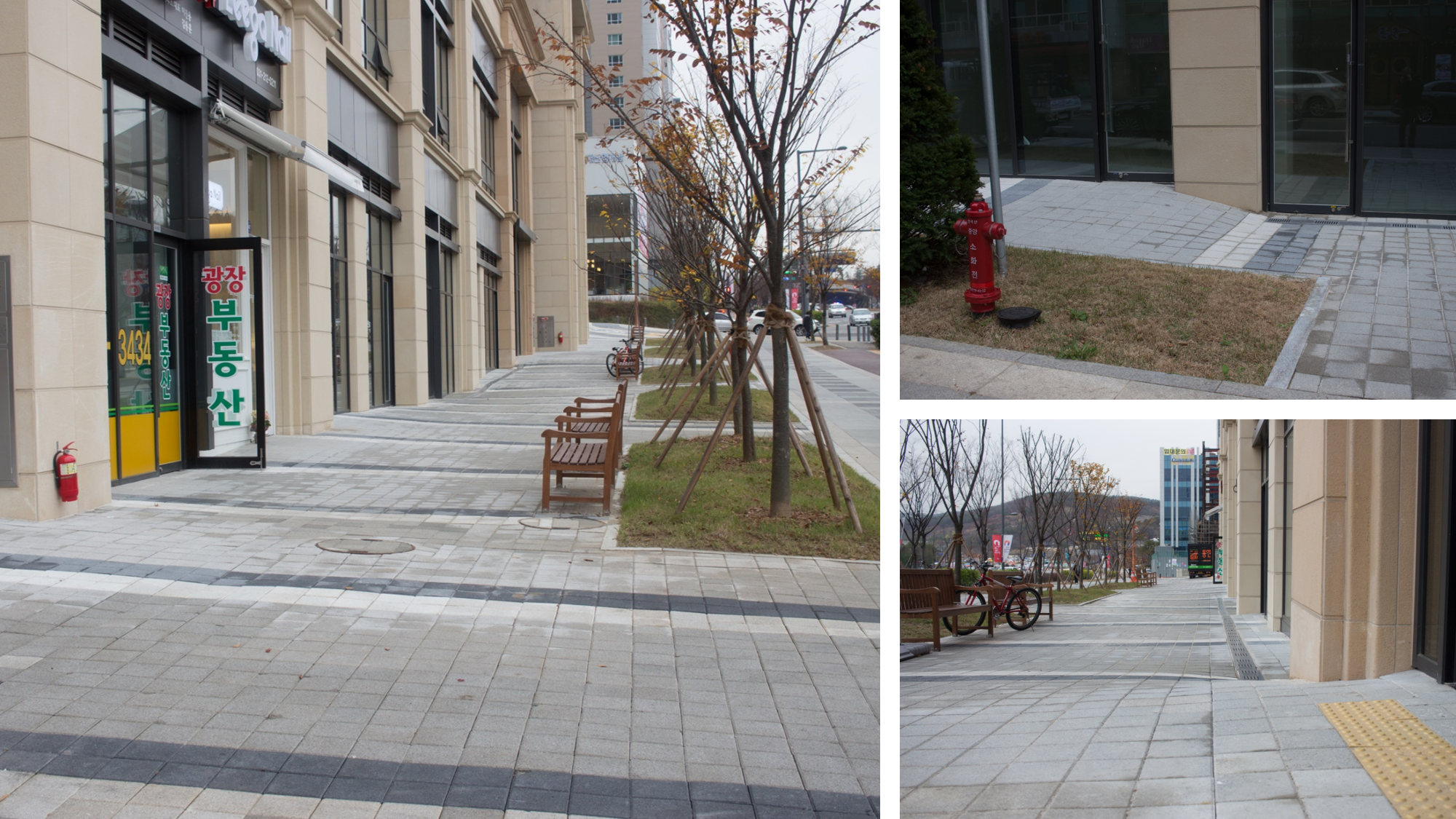}
  \caption{Example of real-world case for varying surface heights. Suwon, South Korea, (Finished Construction in 2015). }
  \label{fig:unevenPhoto}
\end{figure}
Calculating the complexities of human mobility is vital for accurate circulation and time based analysis of paths. When considering mobility with a wheelchair, the option to climb steps is clearly non-existent. However, with a cane or crutches, the question of energy expenditure and time become directly linked to the geometry of the environment. Discussing the impact of stairs on crutch ambulation, ~\cite{gonzalez1994energy} mentions specifically the situation of transportation terminals requiring a 10 m ascent in 200 m horizontal distance requiring 29\% more energy and 45\% more time than a 5\% ramp. When considering locations of the start and end of the steps, as well as the impact different ramp and step gradients would have of these endpoints, the only realistic solution for ensuring human well-being and health is at the forefront of a design is the automated evaluation and feedback to a designer.  

\section{Related Work and Motivation}\label{sec:RelatedWork}
Evaluating a design is a long studied and complex topic. To illustrate the contribution of the proposed method, the following sections break down the problem into two main areas; the data structures that are used to define where an occupant could be in an environment (Sec.~\ref{subsec:navDataStruc} and the methods for evaluating an environment (Sec.~\ref{subsec:evalMethods}). As the method for creating an Accessibility Graph presented in this paper brings together these two, often disconnected components, the literature presented is highly varied in disciplines.  

\subsection{Navigation Data Structures}\label{subsec:navDataStruc}
In robotics, path-planning of an environment for both humanoid and wheeled robots (e.g., autonomous vehicles) is often done in unknown environments such as the DARPA Robotics Competition~\cite{johnson2015team}. In games, human players and autonomous agents must move around a space with directions and paths that are feasible to their goal. Within Architecture, circulation of people moving in the environment and calculating the times it takes to reach various locations are considered. While each of these fields, namely robotics, games, and architecture, is interested in the paths taken within a space, there are also multiple other considerations that impact the approach and use of various techniques. 

The techniques for finding the navigable areas are most prominently different in whether an environmental model exists or not. In robotics, the environment is often unknown and a map must be continuously built for path-planning~\cite{dissanayake2001solution,ge2011simultaneous}. A common method is to create a \textit{height map} -- a grid of cells labeled as a floor or obstacle depending on the height difference between them~\cite{gutmann2005floor}. Similar to the robotics problem of reconstructing real-world environments, in urban analysis a  Digital Elevation Map (DEM)  is created through LIDAR data, identifying areas of roughness using gradient calculations~\cite{putz20163d}. Beyond these techniques for on-the-fly generation, either a polygonal bounds defining a traversable surface is created, or a discretization is performed that represents the space as a graph (nodes and edges). 

\subsubsection{Surface Definition}
In games and character animation, the accessible space of an agent is generated in various ways from a given 3D model. The surface of space designated to an agent's movement is often referred to as the Navigation Mesh~\cite{kallmann2014navigation}. For one, a geometry object can simply be labeled as a traversable surface. In this case, triangulation by subdivision, or a uniform grid can be created, to define links to move across the environment~\cite{narahara2010self}. Two popular approaches to the generation of a navigable space are 1) reconstruction through voxelization~\cite{oliva2013neogen} and 2) mesh vertex calculations~\cite{chazelle1991triangulating,van2011navigation}. Other methods such as multiple horizontal slices with depth-maps~\cite{pettre2005navigation} and prismatic subdivisions~\cite{lamarche2009topoplan} have been shown, with a target of fast agent-based simulation at the expense of pre-processing.

Notably, software such as Unity~\cite{unityHeight} implements an automated generator using voxelization for a \textit{NavMesh} to define surfaces on which a virtual agent can move. However, since architecture models might have thousands of objects and millions of polygons detailing the space (e.g., molding sand decorations), especially objects not related to accessibility (e.g., HVAC systems), this approach is inefficient. While ignoring these unrelated objects is possible, this means the geoemtry is already labeled and the walkable surfaces are known. However, accessibility related issues are still not known even when intended walkable surfaces are defined (e.g., a ramp is a walkable surface but the slope is too steep). 

Alternative methods considered to be true representations thus use vertices of the mesh itself. For generic path-planning of a character the parsing of the environment with little information attributed to the graph is often suitable. For example, the undirected graph created from an Explicit Corridor Map~\cite{geraerts2010planning} or the Implicit Corridor Map~\cite{geraerts2008enhancing} inherently embedded relative obstacle distance information enabling the generation of more realistic paths for an agent to travel along. Improvements on surface-based approaches annotate the environment through the identification of gaps to produce a continuous navigation mesh of the entire walkable region~\cite{vermeulen2018annotating}. 

The focus in these cases is the ability to generate a visually realistic path for an agent to follow in a given environment. In design evaluation, the generation of an agent path for visual realism is less important than the quantification of the space in which that agent would move. It is therefore important to maintain the broad knowledge of where in a model an occupant could physically access, while enabling a data structure that contains location-specific information. 

\subsubsection{Graph and Map Representation}\label{subsubsec:graphRep}

\begin{figure}[pos=h]
  \centering
  \includegraphics[page = 1, width=0.45\textwidth]{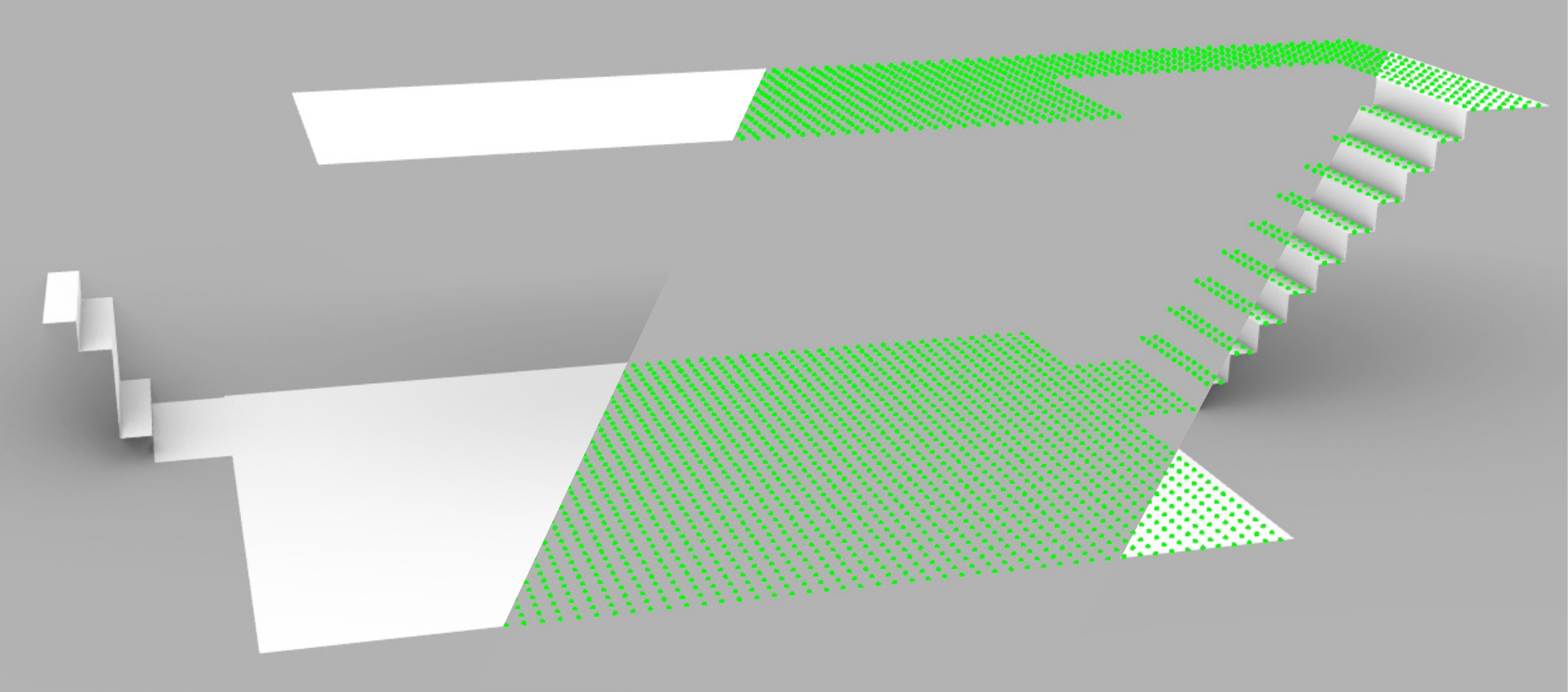}
  \caption{Representation of the environments into a grid of nodes. Illustration showing the geometry, nodes on the geometry for visualization, and the points representing the environment used for analysis in computation. }
  \label{fig:nodeRep}
\end{figure}

A common approach to defining paths and points of interest for spatial analysis involves explicitly defining a network of nodes rather than the construction of a polygon or surface (as seen in a Navigation Mesh). While the polygons of a navigation mesh are able to describe a graph, a dense graph representation of an environment (Fig.~\ref{fig:nodeRep}) has an important difference in usage than general navigation meshes: nodes of the graph are used for additional analyses of the environment. In commonly used analysis methods of the built environment such as lighting, the calculations are reduced to a grid or specific points on the surfaces within the environment(e.g.,~\cite{jakubiec2011diva}). This discrete analysis is particularly useful when considering maximum or minimum values and when it is not easy to know which points are important or of interest ahead of time (e.g., finding the highest intensity of light in a room for glare analysis). Therefore, a dense representation should be created from the start. 

Directed graphs are vital to spatial analysis: the view while going in a straight line from point A to B is 180\textdegree~ from B to A. A visibility graph~\cite{lozano1979algorithm} constructs a graph from the direct connections of all vertices in a set, allowing for optimal shortest path planning; however, the connection to visibility from a human perspective is lacking. This method was used by~\cite{turner2001isovists} to connect a generated grid of nodes along a floor surface which had to be pre-defined by the user. Grid-based graphs have also been constructed in a more localized manner, such that adjacent nodes (or cells) are the only points of contact~\cite{li2010grid,nagy2017buzz}. Similarly,~\cite{nagy2017buzz} used edge culling by checking the intersections of nodes from an even grid on the floor to object geometry. Another approach defines a~\textit{Nav-graph} using a convex hull around vertices of a 3D object projected into a two-dimensional (2D) plane~\cite{doherty2012spatial} .  In particular for path-planning, a dense graph can be created first with weights associated to locations, then construct a hybrid graph after~\cite{kallmann2014navigation, ninomiya2015planning}. 

Often in a design context, minimal, rather than dense graphs are constructed. Sometimes this is due to the input data, such as lines and intersecting points for road analysis~\cite{bielik2012parametric}, or the ability to use BIM data to find doorways~\cite{lee2010computing,lin2018intelligent} to calculate the shortest paths between spaces. When constructing the graph representation of a building, labeled data (e.g., BIM, GIS, IFC) is often used, rather than the geometry in isolation~\cite{lee2010computing,lin2018intelligent,taneja2016algorithms}. Others have used labeled models, but still provide a fallback of surface normals to aid in a tessellation process for defining navigable networks~\cite{boguslawski2016automated}.

While games can rely on reducing model complexity with visual tricks, Architectural models have complex geometries that represent the real infrastructure and that include details often irrelevant to human perception or accessibility\cite{saona1999visibility}. Ray-casting, as implemented for the \ac{AG}, in large and complex scenes can provide an efficient alternative; for example, complex sculptures,  islands, or other scene assets that are visible, but not accessible, are quickly excluded from calculations (e.g., through a bounding volume hierarchy). 

Discussed in Section~\ref{sec:humanFactors}, slope and surface variation is vital information to human mobility, yet methods for considering slope with a graph are limited. An \textit{easiest path} for space syntax was explored in~\cite{nourian2016configraphics} using an abstracted environment representation based on lines (taken from street segments) with a slope defined by the two endpoints.  An indoor topology for human pathfinding which incorporates stairs and ramps was introduced by~\cite{lin2018intelligent}. With a focus on the computational side, research on natural path generation considering slope and terrain variation has been demonstrated in~\cite{RolesSmoothPath, Liu11Shortest}. Another technique used to incorporate 3D geometry is by voxelizing the environment, although these box representations make it difficult to incorporate fine-grained analysis such as slope and varying surfaces~\cite{yuan2010supporting}.While these works in part demonstrate a use of slope to determine routes, the \ac{AG} method is the only one able to combine the dense graph representations without labeled geometry and include multiple factors in the graph. 

\subsection{Common Evaluation Methods}\label{subsec:evalMethods}
Quantifying human experiences in the built environment can be done by either simulating a virtual human (agent-based simulation) or evaluating spatial information represented as a graph (e.g., space syntax, network analysis). These two options are compared and discussed in the following sections. 

\subsubsection{Agent Based Analysis}

A common approach to evaluating the way occupants would utilize a given design is to virtually recreate the individual occupants, referred to as agents. Agents are assigned some tasks and utilize the navigation data structures described in Sec.~\ref{subsec:navDataStruc} to provide feedback on how those agents experienced the space. Likewise, the collective movement of the agents based on those data structures can evaluate the efficacy of the design. Most similar to game-based evaluation,~\cite{narahara2007space} implemented a responsive narrative in which a character moves and interacts with the environment, offering subjective feedback through visualization. Through multiple iterations of agents moving in the environment, statistics and data are extracted about how the agent performed and experienced the space. Influential work for evaluating an architectural space was in the application of egress~\cite{rahman2008agent,gwynne1999review}. In simulating the probabilistic movement of a crowd during evacuation, social and behavioural factors have significant impact on the accuracy~\cite{pan2007multi}. For egress, the value for analysis is relatively straightforward, i.e., length of time for occupants to exit. When deriving useful data on the complex experience of an occupant, the inputs to the simulation greatly increase. 

In addition, in a narrative-based model for multi-agent simulation,~\cite{schaumann2019simulating} created a simulation around spaces, actors, activities, and narratives. The data extracted is of walking paths, density, length of stay, and social interactions. As even the same person does not act identically on multiple days, numerous iterations of agents must be simulated to demonstrate a convergence towards accurate data. The challenge in this approach is the efficacy of properly representing aspects such as the social interaction between agents. In true human interactions, aspects as small as eye contact can change the social dynamic and experience~\cite{lang2012better}. Through deep learning, ~\cite{nakada2018deep} developed a model to control a virtual biomechanical human to the point of photoreceptor responses that inform head directions and gaze. While agent based simulation addresses an important aspect of time, the difficulty in modeling true social interactions suggests, at least for the time being, that more concrete metrics in the environment independent of social factors must be included in spatial evaluation. For example, as mentioned previously, the curvature of a surface directly affects human occupants as they traverse it, regardless of the point in time. 

\subsubsection{Graph Based Analysis}
In a more discrete approach than agent-based simulation, graph techniques work towards integrating known numeric quantities about a space, or attributes of locations in the space, to a graph. While in Section~\ref{subsubsec:graphRep} the specific methods of~\textit{creating a graph} were compared with the current work, this section will discuss the evaluation methods and criteria for~\textit{analyzing a graph} in design. Network analysis in the built environment, in its simplest form, assesses relationships between various locations. Such spatially relevant networks consisting of nodes and edges have been demonstrated in BIM~\cite{lee2010computing,suter2013structure}, but are difficult to construct with unlabeled data. By using intersections and properties of the environment (3D model), networked relationships of a space can be constructed. Early work in automatic evaluation was in the use of polygons representing a field of view (ISOVISTS)~\cite{benedikt1979take}. The intersecting polygons can then be reduced to numeric values of visibility intensity for a given location, which was later modified to a graph-specific data structure~\cite{turner2001isovists}. More recently,~\cite{varoudis2014beyond} implemented an all-to-all connection of nodes to construct a 3D visibility graph for space syntax analysis. These analysis methods are all possible with the \ac{AG}, while the 3D visibility graph can be created with human-accessible locations by offsetting the \ac{AG} vertically in some discrete increments. 

While network and circulation analysis often uses the shortest path algorithm, the physically shortest path is not necessarily a good metric of the way in which people will move within the environment nor the experiences an occupant would have. With metrics such as~\textit{walkability} (referred to in Section~\ref{sec:intro}) commonly using the path length to represent a unit of time,~\cite{nourian2016configraphics} reduced multiple metrics to a single value of time of travel for analysis. However, the relationship between speed, energy, and human ability is well researched in the literature provided in Sec.\ref{sec:humanFactors} and reducing to a time-based metric is often not representative. In the case of accessibility, the physically shortest path may not be the physically least demanding nor the one an occupant may take. The types of data used in graph-based analysis of paths are wide and varying, such as calculating the number of turns along an occupant path~\cite{Fuchkina2017,shin2019indoor,nourian2016configraphics, lee2018circulation}. Similarly, the types of users have a profound impact on the importance of different factors when considering accessibility, such as considering direct connectivity and spatial depth for elderly housing~\cite{lee2018circulation}. 

One challenge in a sparse network can be seen in~\cite{shin2019indoor}, where an Indoor Walkability (IW) factor was used to quantify paths based on various weights.  As the graph used in path generation consists of sparse nodes connecting maximal spaces, the weighted values are path-based, rather than information on what is~\textit{along} the path. The authors proposed future work, including object-level refinement. However, it is unclear how the sparse network of locations used in the UCN~\cite{lee2010computing} would function in this context. For example, while a table has geometry that can be included in the network, overhead lighting casting a bright spot in the middle of a room would be ignored. While the motivation of~\cite{shin2019indoor} is similar to the \ac{AG}, a key difference is the scalability and limitations related to both requiring labeled data and limited nodes throughout the environment. 

While these past studies have varying resolution to their graph structures, they share a common theme, emphasizing the utility of graph theory in environmental analysis. However, the focus in the past works of space syntax and network analysis has been primarily on the relationship between the environment and building rules \& regulations, while the work presented here is based on the direct human factors involved--with the motivation to influence the rules \& regulations devised around these human experiences. Additionally, the parameter objects described in ~\cite{shin2019indoor,lee2018circulation} could be extended using the graph information generated within this paper and categorize a given path as listed in Table~\ref{table:surf_type}.

\section{Accessibility Graph Method}~\label{sec:methodology}
\begin{figure}
  \centering
  \includegraphics[page = 16, width=0.48\textwidth]{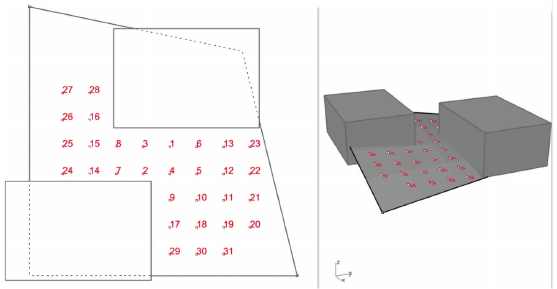}
  \caption{The breadth-first process of defining node locations. The numbers correspond to the order the node was checked. Two box objects are defined as obstacles while the surface itself is a free form shape in 3D. Further examples are given in the Appendix and Supplemental Materials.  }
  \label{fig:bfs}
\end{figure}

The resulting graph is a mapping of graph vertices to point locations in the environment, referred to as nodes. First, an interpretation of the 3D model must be done in order to find the nodes accessible from a starting point (Fig.~\ref{fig:bfs}). Once this mapping has been created, for each node in the environment, any human based analysis can be performed and attributed to that location. Considering the nodes in isolation, a visibility graph can be generated~\cite{turner2001isovists}. More complex and individual factors can also be analyzed, such as falling injury probability~\cite{Schwartz:2018:HTD:3289750.3289768}. 

As with any building-based graph, the shortest path between two nodes can be found. When generating a graph based on a specific metric, such as visibility, information about the access to a node is lost. This is exemplified in the problem of an environment location not being accessible at all, while still calculating the values associated with those locations. The approach introduced in this paper considers the method of movement from one node to the next, allowing for a quick end to the computation when starting from a poorly accessible location\footnote{This can be seen in Appendix~\ref{appendix} and the supplemental material when a ramp is too steep or steps are not useable.}. Combining the movement from one node to another with the node based analysis, numerous considerations for generating a likely or optimal path can be used. The factors of one directed edge cost can be weighted against others, as well as any attribute associated with a node. Furthermore, the movement type can be used in path planning and analysis, where the \textit{cost} of a path for a search algorithm (e.g., Dijkstra~\cite{dijkstra1959note}) is built on weights such as slope. 

The method presented expands the ray-casting based breadth-first search to find accessible locations of an environment given an initial starting position as described in~\cite{Schwartz:2019:nonflat}. In this section, the method of the graph generation is explained followed by example implementations of both path planning and path analysis. 

\subsection{Graph Components}\label{sec:graphComponents}

\subsubsection{Graph Definition}\label{sec:graphDef}

To define the graph of an environment, a starting point $\tau \in \mathbb{R}^3$ within the CAD environment is defined from which a breadth-first process begins. The graph $D := (V,E,W)$ is a weighted directed graph (digraph) composed of the set of vertices $V$, edges $E$, and a set of weight vectors $W$. The set of vertices is defined as:

\begin{rmk}
$V(D) := \{v_1,\dots,v_n \}$
\end{rmk}

The edge set of $D$ is defined as:

\begin{rmk}
$E(D) := \{e_1, \dots,e_k ~|~ e_{ij} = (v_i,v_j) \land v_i, v_j \in V(D) \} $
\end{rmk}

The last basic component of the digraph is the set of weight vectors:

\begin{rmk}
$W(D) := \{ \mathbf{w_1}, \dots, \mathbf{w_k} \}$
\end{rmk}

Where $\mathbf{w}$ is a vector of weights with a mapping to edges such that $\Psi : e_{ij} \rightarrow \mathbf{w_{ij}} $. The weight vector $\mathbf{w_{ij}}$ contains the different costs associated with the movement on an edge. A subscript is used to denote a specific weight in the vector, such that the weight corresponding to distance is denoted with subscript $d$, such that $\wm{d} =  \texttt{dist}(v_i,v_j)$. \\ 

To map the graph to a spatially relevant set of locations, the set of vertices have an intrinsic mapping to 3-dimensional points in the environment. 

\begin{rmk}
 $v_i \mapsto n_i  | v_i \in V \land n_i \in \mathbb{R}^3$
\end{rmk}

In the following sections, a reference to $n_i$ is to reference the 3-dimensional point of $v_i ~\in V(D)$, and similarly, a line segment $\mu_{ij} \mapsto e_{ij} | \mu_{ij} = \overline{n_in_j}$.

A set of geometry objects $\Gamma$ represents the interactable surfaces relating to human accessibility.  

\begin{rmk}
 $\Gamma := \{ B_i \dots B_{i+1} \}$
\end{rmk}

The set $B_i$ represents a collection of geometry objects with a particular property. For any generated graph $D$, $ |\Gamma| > 0$ for at least one traversable surface must be defined (all objects can default to traversable). Additional surfaces can include obstacle objects as well as specific surface types. For example, using BIM: hardwood or carpet flooring could be defined, or with GIS: sidewalk concrete and grass could be included.

\subsubsection{Subgraph and Subsets of Locations}
\begin{figure}[pos=h]
  \centering
  \includegraphics[page = 6, width=0.45\textwidth]{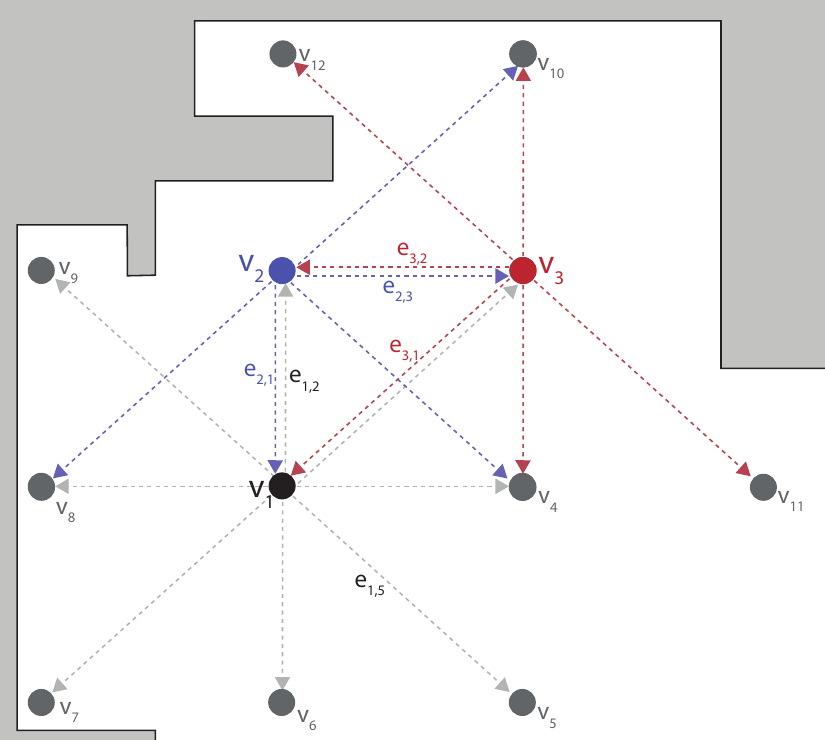}
  \caption{The relationship of vertices and edges in the $(xy)$ plane. Each color group (grey, blue, and red) represent the edge induced subgraph \edgeSubgraph. The parent-child relationship is defined as the from and to nodes, e.g., $v_1$ is the parent and $v_5$ is a child of $v_1$. }
  \label{fig:parent-child}
\end{figure}

When discussing the graph in terms of accessibility, the basic concept is given as: from a given node, what is the cost, if possible, to get to the next node? The graph is built such that the only nodes to move to, from any particular node, are ones that were checked specifically in relation to that node, which requires a unique check for each time a node is defined to have an accessible path to another node (Fig.~\ref{fig:parent-child} uses colored edges to illustrate connections between nodes). In other words, a parent node is the starting (initial) node ($n_1$) of a directed edge of an ordered pair of nodes $(n_1,n_2)$ and the child is the ending (terminal) node ($n_2$). This ordered pair (parent,child) is later abbreviated as $(p,c)$. However, discussed in the following sections, the use of neighboring nodes defines edge costs for cross-slope costs. 

Without additional processing of the graph, it may be weakly connected, as a graph may have a directed edge to a node that has no outgoing edges. This may happen if the \ac{AG} is generated with areas that allow an occupant to get to a location but not back (e.g., a ramp that is feasible to traverse down, but not up). A separate issue is when a node is defined as a child through the given parameters and subsequently has no children. For this case, a specific parent set $P \subseteq V$ of nodes is defined.

\begin{rmk}
$P := \{v_i | (v_i,v_j) \in E \} $ 
\end{rmk}

As each node in $P$ has an outgoing edge, there is at least one child node associated with it. The neighbourhood $N^+_D(v_i)$ is the set of vertices dominated by vertex $v_i$ in graph $D$. While dense graphs are sometimes scrutinized for the space complexity, this is most often seen in fully connected visibility graphs and not applicable here as edge connections are limited to a predefined number of immediate neighbours (in euclidean space) of $v$, creating a sparse graph. Figure~\ref{fig:parent-child} shows the base case in which only adjacent points on the grid are possible child nodes (an extension of this is described in the following sections).

For clarity, $N^+_D(v_i)$ is shortened and referred to as the child set $C$ when the context is of the general node $v$. Next, a set of outgoing edges from a vertex $v_i$ is defined as:

\begin{rmk}
$ E^{\rightarrow}_D(v_i) := \{ e_{ij} | ~v_j \in N^+_D(v_i) \}$
\end{rmk}

This set of outgoing edges is then used to define the edge-induced subgraph \edgeSubgraph of $v_i$, which is seen in Figure~\ref{fig:parent-child} through different colorings. 

As $\forall v_i \in P,~\exists ~\edgeSubgraph$, the weights associated with an edge $e_{ij} \in \edgeSubgraph$ can be influenced by both node and edge values of its subgraph so that human-relevant factors such as cross slope can be accounted for.

\subsubsection{Graph Creation Parameters}\label{sec:graph_params}
\begin{figure}[pos=h]
  \centering
    \includegraphics[page = 6, width=1in]{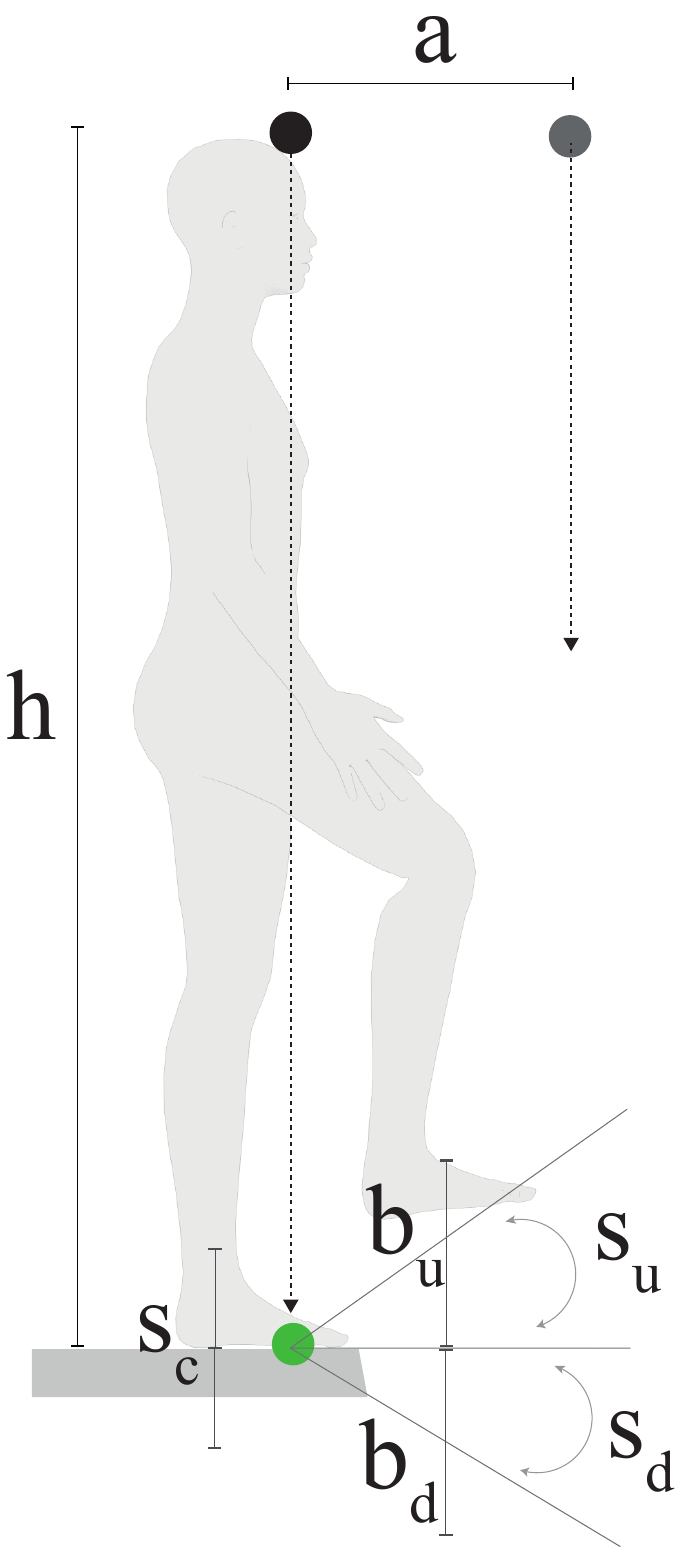}
  \caption{The parameters for traversing the model during the graph creation. Cross-slope $s_c$ is in the perpendicular plane to the figure, as illustrated in Figure~\ref{fig:crossSlope}).}
  \label{fig:params}
\end{figure}

The graph generation results in the defined locations of accessibility given specific parameters.  These parameters are shared among many of the navigation mesh algorithms (e.g.,~\cite{unityHeight}) and while defaults can be used, are input by a user.  Parameters used within this algorithm are:
\begin{enumerate}[]
    \item $\tau$ : Starting Point (x,y,z)
    \item $h$ : Height
    \item $a$ : Grid Spacing (x,y)
    \item $b_u$,$b_d$ : Step Height (Ascent +, Descent -)
    \item $s_u$,$s_d$,$s_c$ : Slope (Positive +, Negative -, Cross)
    \item $\gamma$ : Minimum Required Children
    \item $\Phi$ : Set of Neighboring Directions
\end{enumerate}

Figure~\ref{fig:params} illustrates the decision parameters from a given node (shown in green) to another. The slope and step parameters are calculated separately. Slope is defined as an increase or decrease of height from one node to another when no geometry obstructs a direct path between nodes. A step is when an obstruction of the path between nodes exist. Uniquely, $\gamma$ is the minimum children required to consider a node valid, and $\Phi$ are the direction vectors to check for valid children such that a set of 8 directions (0 and $\pm$1) would be the adjacent 8 nodes while another direction set (e.g., 0, $\pm$1, $\pm$2) may include a secondary level of subsequent nodes. Given 8 neighboring nodes, a $\gamma$ value of 8 would mean a node is only valid in the graph if it is also able to access all directions. This check would fail if a node is close to a wall, creating a minimum buffer region. 

These components allow for a great variety of human related accessibility issues to be incorporated in the graph generation. For example, \textit{slope} can be defined as an incline of 12:1 while the minimal \textit{step height} can be 0 cm. In this case, there is a rule describing an allowable height difference when traveling between two locations; as long as there is a direct connection between those locations (i.e., it is a ramp or incline, not an obstacle or step), describing the graph required of a wheelchair. Parameters can be set at some maximal human capable level (maximal clearance) with the values stored within multiple graphs as edge costs and later iterated through for various path-finding and accessibility checks. Alternatively, a single graph can be built with the edge costs incorporated on-the-fly. As a note, post-calculating edge costs is embarrassingly parallel as this is done $\forall ~v_i \in P$, where $~\exists ~\edgeSubgraph$.

\subsection{Algorithm Overview}

\begin{figure}[pos=h]
  \centering
  \includegraphics[page = 3, width=0.48\textwidth]{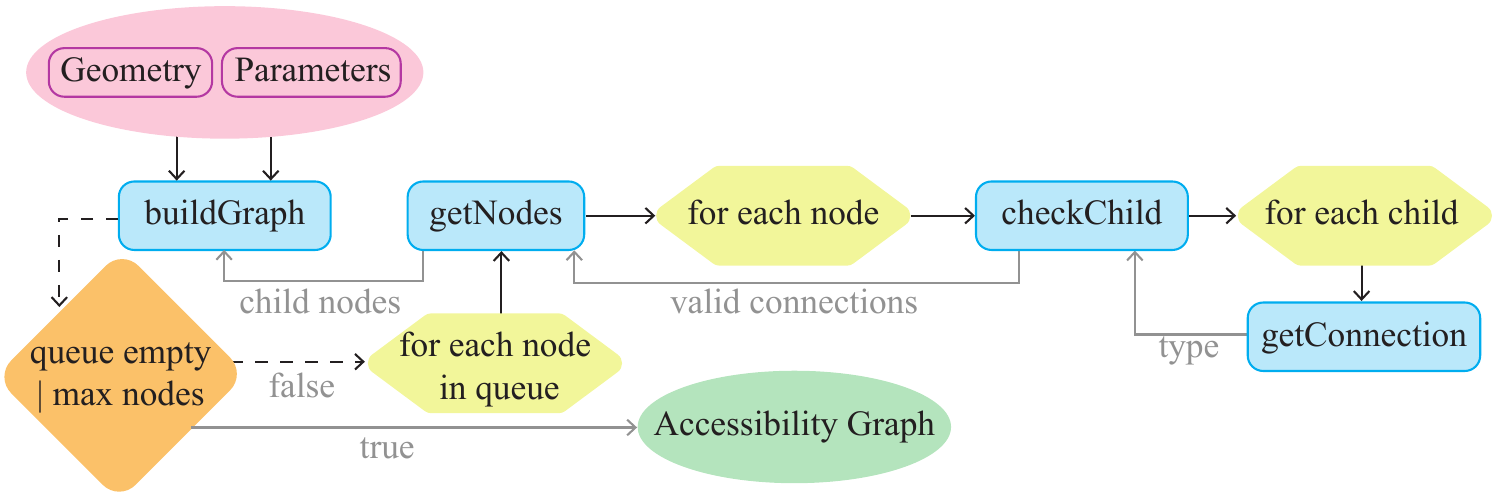}
  \caption{Overview of the algorithm to generate the accessibility-based graph. Blue boxes are the methods detailed in this section. }
  \label{fig:AlgOverview}
\end{figure}

In Fig.~\ref{fig:AlgOverview}, a simple overview of the algorithm to build the accessibility-based graph is shown with four main functions described in the following sections. The algorithm initializes with the user parameters and geometry objects that remain in a global scope. Overall, the algorithm iterates through a queue of nodes that are continually added to the queue in a breadth-first manner from the starting location $\tau$ within the environment (\texttt{buildGraph}). While checking each item of the queue, \texttt{getNodes} iteratively finds the valid neighbor nodes by calling \texttt{checkChild} on each potential neighbor. The final function in the process, \texttt{getConnection}, determines if an edge can exist between two nodes based on the input parameters, and if so, the type of movement required between the original and potential child node. Once an edge is determined to be possible, the movement type is returned (likewise for all connections for the original queued item). Before the breadth-first process continues to the next item in the queue, the graph is updated with vertices, edges, and weights.

\subsubsection{Breadth-first Process}

The generation of common surface and graph information methods regarding navigation takes into account the entire environment at once. Breadth-first search allows for the fast completion of the graph generation when no accessible locations are found (\nameref{appendix}). This feature is particularly useful in the planning and design phase of a building or environment as it offers immediate numerical and visual feedback about the features that are blocking accessibility.  The quick ending to the graph generation benefits the calculation time in practical use cases (e.g., the starting location being a ramp that is too steep). 

Generating the graph begins with a valid starting location where start point $\tau$ is above a valid surface geometry in $\Gamma$. The algorithm begins with an empty graph $D$ and an ordered set $Q$ representing the queue of vertices that remain to be checked for validity in $P$. At initialization $Q = \{\tau\}$, with the algorithm completing when $Q = \varnothing$. 
\begin{algorithm}
 \SetAlgoNoEnd
 \DontPrintSemicolon
 \SetKwFunction{FMain}{buildGraph}
 \SetKwProg{Fn}{Function}{:}{}
 \SetKwFunction{Finner}{getNodes}
  \SetKwFunction{Fnn}{setEdgeCost}
  \SetKwFunction{Fnnn}{buildGraph}
 \SetKwProg{Fni}{}{}{}
 $Q \gets \tau$ \;
 $V \gets \tau $ \;
 $E,W \gets \varnothing,\varnothing $ \;
 $D \gets \{V,E,W\}$ \;
  
 \Fn{\FMain{$D,Q$}}{ 
    \If{$Q = \varnothing$}{
    \bf{return}~ $D$\;
    }
    $p \gets Q.pop()$ \; \label{algln:pop}
    $ C,w_t \gets$ \Fni{\Finner{p}}{}
    $W \gets w_t $\; \label{algln:setSteptype}
 \For{c $\in$ $ C $}{
    $ W \gets$ \FuncSty{setEdgeCost($p,c$)} \; \label{algln:edgecost}
    $ V \gets c$ \;
    $ E \gets (p,c) $\;
        \If{$c \not\in P$}{
        $Q.insert(c)$\; 
        } 
 }
   \bf{return}~ \Fni{\Fnnn{D,Q}}{}
 }
 \caption{Build digraph of nodes and edges}
 \label{alg:buildgraph}
\end{algorithm}

Generating the graph can be seen in Algorithm ~\ref{alg:buildgraph} line~\ref{algln:pop}, where a parent node $p$ is removed from the queue $Q$. The valid children nodes are assigned to set $C$ and if a child node is not already defined as a parent node, it is added to $Q$. Once the set of valid children are returned, the weight vector containing step type is stored in the weight set (line~\ref{algln:setSteptype}). This pseudocode shows the method in which weights are calculated during graph construction, and therefore, during the iteration over the child set, a function \FuncSty{setEdgeCost()} on line~\ref{algln:edgecost} calculates a directional edge cost between the two nodes and stores it in $W$. This function can be used to set multiple types of edge costs, such as distance, slope, and energy, as discussed later. After the directed edge of a parent and child is stored, the child node is checked for existence in the parent set. If the node is not in the existing parent set it is added to $Q$.

Two main functions are defined that determine the validity of a child node within the \FuncSty{getNodes()} function (called in Alg.~\ref{alg:buildgraph} and later defined in Alg.~\ref{alg:getNodes}). The function \FuncSty{inter($n_i,n_j$)} is defined as returning the distance $\overrightarrow{{n_in_j}}$ intersects $B_i$ in the environment and \FuncSty{dst($n_i,n_j$)} returns $\| {\vec{n_i}-\vec{n_j}}\|$. 

\begin{figure*}
\centering
    \begin{subfigure}[b]{0.325\textwidth}
        \includegraphics[page=4, width=\textwidth]{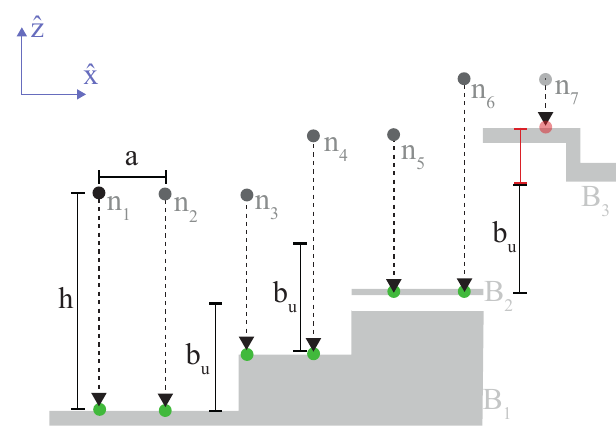}
        \caption{Node position above step height}
        \label{fig:heightParsing_1}
    \end{subfigure}
    \begin{subfigure}[b]{0.325\textwidth}
    	\includegraphics[page=2, width=\textwidth]{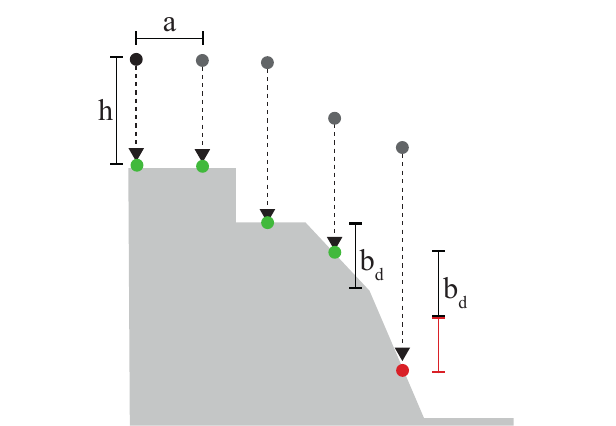}
    	\caption{Node position below step limit}
    	\label{fig:heightParsing_2}
	\end{subfigure}
    \begin{subfigure}[b]{0.325\textwidth}
    	\includegraphics[page=3, width=\textwidth]{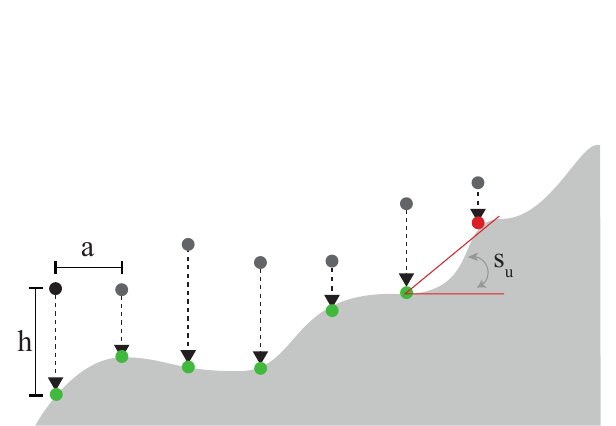}
    	\caption{Node position outside slope limit}
    	\label{fig:heightParsing_3}
	\end{subfigure}
    \caption{Illustration of three common occurrences in graph creation of non-flat surfaces for checking a valid child node. Green circles represent a valid node and red circle as invalid based on the parameters given. The black circle represents the starting node $n_1$ and each grey circle is a subsequent check for a valid node. ~\ref{fig:heightParsing_1} represents climbing stairs and multiple geometry objects, ~\ref{fig:heightParsing_2} is descending stairs, and ~\ref{fig:heightParsing_3} is a natural topography. The letters $a,h,b$ represent the offsets used as values in the graph creation. ~\ref{fig:heightParsing_1} shows the node numbering scheme with the same ordering followed in ~\ref{fig:heightParsing_2} and ~\ref{fig:heightParsing_3}.} 
    \label{fig:heightParsing} 
\end{figure*}

Figure~\ref{fig:heightParsing} shows the node evaluation in the $(xz)$ plane where the ray cast direction is $-\hat{z}$. As the graph is in relation to physical space, $-\hat{z}$ corresponds to the direction of gravity. The parameters and labels used are those described in~\ref{sec:graph_params}. The surfaces illustrated provide examples of a step, faceted surface, and natural topography (e.g., a hill or mountain). When a possible node is invalid, it is not added to $Q$, and in the simple example within the figure, the graph completes. To simplify the illustration, Figure~\ref{fig:heightParsing} is a two-dimensional example is given and only the directed relation from the left to right is discussed. 

In \ref{fig:heightParsing_1}, a ray is cast in $-\hat{z}$ starting at $n_1$. As the node found from \FuncSty{inter($n_3,n_3-\hat{z}$)} $< b_u$, $n_3$ is considered valid. When checking $n_4$, $h$ is reset based on the previous node ($n_3$). To demonstrate the use of $\Gamma$, three objects, $\Gamma = \{B_1,B_2,B_3\}$ are represented in the illustration where they are all walkable surfaces. The last node, $n_7$, is shown in red as \FuncSty{inter($n_7,n_7-\hat{z}$)} $> b_u$, ending the graph generation. In the down step case for parameter $b_d$, \ref{fig:heightParsing_2} illustrates the intersection point for $n_5$ falling outside the threshold. Finally a node $n_7$ is invalid in \ref{fig:heightParsing_3} as the angle created from $n_6$ and \FuncSty{inter($n_7,n_7-\hat{z}$)} $> s_u$. Furthermore, the relative positioning in $Z$ of parent-child nodes can be seen by the height variation relative to the previous node.  

Algorithm~\ref{alg:getNodes} implements this process in three-dimensions. The function \FuncSty{getNodes(p)} takes an input node $p$ and tests for its children in the directions contained in a given set of an even grid in the $(x,y)$ plane (i.e., Fig.~\ref{fig:parent-child}), represented in the set $\Phi = \{-1,0,1\} \times \{-1,0,1\}$. With these directions, a $45\deg$ angle is a limiting factor for a path connecting nodes. To decrease this angle, $\Phi$ can incorporate additional locations, i.e., (-2,-1), which would correspond to node $v_3$ connecting to $v_8$ in Figure~\ref{fig:parent-child}. An alternative approach to expanding the neighboring nodes could be to iteratively expand while some condition remains true, such as no stepping, through an adaptive highway graph~\cite{ninomiya2015planning}. Algorithm~\ref{alg:buildgraph} and~\ref{alg:getNodes} describe the three-dimensional breadth-first process in general terms. On line~\ref{algln:pxyz}, a point $c$ is composed of the $x,y,z$ values of $p$, offset in the directions of $\Phi$. 
 
\begin{algorithm}
 \SetAlgoNoEnd
 \DontPrintSemicolon
 \SetKwFunction{FMain}{getNodes}
 \SetKwProg{Fn}{Function}{:}{}
 \SetKwFunction{Finner}{checkChild}
 \SetKwProg{Fni}{}{}{}
 \Fn{\FMain{$p$}}{
\tcp{Check parent $p$ for valid children}
 $C,w_t  \gets \varnothing,\varnothing$ \;
 \For{$(i,j) \in \Phi$}{
        c $\gets$ ($p_x+(i\times a),p_y+(j\times a),p_z + h$)\; \label{algln:pxyz}
         $t,c\prime \gets$ \Fni{\Finner{p,c}}{}
         \If{$c\prime$}{
          $C_{ij} \gets c\prime$\;
          $w_{t_{ij}} \gets t $\;
         }
  	 }
 \If{|C|<$\gamma$}{
  \bf{return} $\varnothing,\varnothing$\; \label{algln:returnEmpty}
 }
 \Else{
  \bf{return} $C,w_t$\; \label{algln:returnC}
 }
 }
 \caption{Check for valid neighbors}
 \label{alg:getNodes}
\end{algorithm}

The process for determining the return value of \FuncSty{checkChild(p,c)} begins with the description of Figure~\ref{fig:heightParsing} and is expanded to define edge weights and determine criteria for an edge connection in the following section.

\subsubsection{Determining Accessibility}\label{sec:graph}

As discussed in the previous section, each $(p,c)$ is checked against a set of parameters and criteria. $N^+_D(p)$ is constructed in Alg.~\ref{alg:getNodes}, represented by the returned set $C$. One aspect of accessibility is a minimum width or region a valid path can traverse through. Algorithm~\ref{alg:getNodes} can be expanded to force accessible edges in path-finding to only pass through nodes that are connected to at least some user-defined number of neighbors. On line~\ref{algln:returnC} of Alg.~\ref{alg:getNodes}, rather than returning $C$, a check can be done to see if $|C|$ is greater than some minimum connectivity and if not, return an empty set. For example, if $\Phi$ is the set of 8 neighboring directions, \texttt{return} $C \iff |C| = 8$.

For $c \in C$, $c$ is determined to be valid by function~\FuncSty{checkChild(p,c)}, shown in Alg.~\ref{alg:checkChild}. First, the distance to the closest $B_i \in \Gamma$ is found by \FuncSty{inter($c,c-~\hat{z}$)} . Given no labeled geometry in $\Gamma$, $B_w$ is the same set of objects as $\Gamma$. For labeled geometry consisting of walkable and non-walkable surfaces, a set of walkable surfaces are defined as $B_w \subset \mathbb{R}^3$. If the closest distance is to $B_i \not \in B_w$ the node is not valid. The connection type $t$ between $p$ and $c^\prime$ is calculated by \FuncSty{getConnection(p,$c^\prime)$} and if valid, the step and slope parameters are verified. In implementation, the relational information of connection types between the parent and child nodes is stored during the graph construction, and the function returns the connection type ($t$) and valid node ($c^\prime$). 

\begin{algorithm}
 \SetAlgoNoEnd
 \DontPrintSemicolon
 \SetKwFunction{FMain}{checkChild}
 \SetKwProg{Fn}{Function}{:}{}
 \SetKwFunction{Finner}{getConnection}
 \SetKwProg{Fni}{}{}{}
 \SetKwFunction{Finter}{inter}
 \Fn{\FMain{$p,c$}}{
\tcp{Check if child $c$ is a valid child}
    $c^\prime \gets $\Finter{$c,\langle c_x,c_y,c_z-1 \rangle $}\;
    \If{$c^\prime \not \in B_w$}{
    \bf{return} $\varnothing,\varnothing$\;
    }
    $t \gets$ \Fni{\Finner{p,$c^\prime$}}{} \label{algln:callGetConnect}
    \If{$t \not = t_{\texttt{INVALID}}$}{
        \If{$t = t_{\texttt{DIRECT}}$}{
            \If{$ s_d < (p_z - c^\prime_z) < s_u$}{
                \bf{return} $t,c^\prime$\;
            }
        }
        \If{$ b_d < (p_z - c^\prime_z) < b_u$}{
            \bf{return} $t,c^\prime$\; 
        }
    }
}
 \caption{Check if a child is within constraints}
 \label{alg:checkChild}
\end{algorithm}

\begin{figure*}
\centering
    \begin{subfigure}[b]{0.19\textwidth}
    	\includegraphics[page=2, width=\textwidth]{images/Connections.pdf}
    	\caption{Direct Connection}
    	\label{fig:connectionRay1}
	\end{subfigure}
    \begin{subfigure}[b]{0.19\textwidth}
    	\includegraphics[page=5, width=\textwidth]{images/Connections.pdf}
    	\caption{Step Over}
    	\label{fig:connectionRay2}
	\end{subfigure}
    \begin{subfigure}[b]{0.19\textwidth}
        \includegraphics[page=1, width=\textwidth]{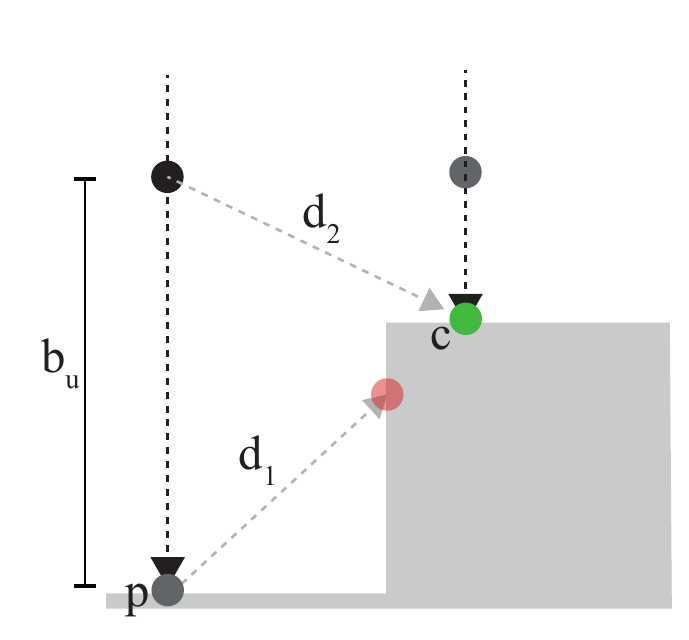}
        \caption{Up Step}
        \label{fig:connectionRay3}
    \end{subfigure}
    \begin{subfigure}[b]{0.19\textwidth}
    	\includegraphics[page=3, width=\textwidth]{images/Connections.pdf}
    	\caption{Down Step}
    	\label{fig:connectionRay4}
	\end{subfigure}
    \begin{subfigure}[b]{0.19\textwidth}
    	\includegraphics[page=4, width=\textwidth]{images/Connections.pdf}
    	\caption{Not connected}
    	\label{fig:connectionRay5}
	\end{subfigure}
    \caption{Example configurations for a connection between nodes. First, a check $d_1$ is performed (\ref{fig:connectionRay1}). When the first check is not successful, depending on the relation between the nodes, either the up(\ref{fig:connectionRay3}) or down step(\ref{fig:connectionRay4}) limits are used for a second check $d_2$. If $c_z = p_z$, a special case is defined(\ref{fig:connectionRay2}). In~\ref{fig:connectionRay5}, the incline after the parent and down step of the child block a connection. This last case is the same result given a wall or another obstacle preventing an edge between the nodes.} 
    \label{fig:connectionRay} 
\end{figure*}

The type of connection between two nodes is determined as one of five possible cases (Fig.~\ref{fig:connectionRay}): No Step~\ref{fig:connectionRay1}, Step Over~\ref{fig:connectionRay2}, Step Up~\ref{fig:connectionRay3}, Step Down~\ref{fig:connectionRay4}, Invalid~\ref{fig:connectionRay5}. The reference to "No Step" is specifically regarding a geometry that protrudes off of the ground surface, which would require a foot, wheelchair, or other aide to lose contact with the ground during traversal. The "No Step" is referred to as a direct connection, which works in both flat and inclined ground as the height offset between nodes is also stored. Therefore, future queries of the relationship between nodes (such as defining edge cost) returning both a height offset and direct connection allow for a specific categorization such as an incline instead of a step. This method provides a similar but more detailed approach to determining node connections as the edge culling in~\cite{nagy2017buzz} or the edge connectivity determination of a visibility graph~\cite{turner2001isovists,Ghosh2013}.

Given the first connection type being false, a second ray, referred to in the figure as $d_2$, is calculated and used to check for a connection. Depending on the height relationship between $p$ and $c$, the translated start position of the ray is determined by $b_u$ or $b_d$ as the directed edge is either stepping up or down. The special case is for when the two nodes are equal in height with an obstruction between them, requiring a step over the obstruction (Fig.~\ref{fig:connectionRay2}), which has a ray start point calculated the same as a step up. 

\begin{algorithm}
 \SetAlgoNoEnd
 \DontPrintSemicolon
 \SetKwFunction{FMain}{getConnection}
 \SetKwProg{Fn}{Function}{:}{}
 \SetKwFunction{Finter}{inter}
 \SetKwFunction{Fdst}{dst}
 \SetKwProg{Fni}{}{}{}
 \Fn{\FMain{$p,c$}}{
\tcp{Check parent $p$ for connection type to $c$}
     \If{\Fdst{$p$,$c$} $\leq$ \Finter{$p$,$c$}  }{
      \bf{return} $t_{\texttt{DIRECT}}$\; \label{algln:direct}
     }
     \If{$p_z \leq c_z$}{
     $p_z \mathrel{{+}{=}}b_u$\;
         \If{\Fdst{$p$,$c$} $\leq$ \Finter{$p$,$c$}  }{
         \If{ $p_z = c_z$ }{\bf{return} $t_{\texttt{OVER}}$}
         \bf{return} $t_{\texttt{UP}}$
         }
     }
     \If{$p_z > c_z$}{
     $p_z \mathrel{{+}{=}}b_d$\;
         \If{\Fdst{$p$,$c$} $\leq$ \Finter{$p$,$c$}  }{
          \bf{return} $t_{\texttt{DOWN}}$\;
         }
     }
\bf{return} $t_{\texttt{INVALID}}$\;     
 }

 \caption{Check for node connection type}
 \label{alg:getConnection}
\end{algorithm}

Algorithm~\ref{alg:getConnection} demonstrates the process for calculating the rays and connection types. After the connection type and parameters of $C$ is determined, the algorithm returns to \FuncSty{buildGraph(D,Q)} on line~\ref{algln:callGetConnect}. At that point, for the given $p$ passed to \FuncSty{getNodes}, the subgraph of $p$ can be constructed ($D[E^+_D(p)]$).     

\subsection{Edge Costs}\label{sec:edgeCosts} 

The cost of traveling from one node to another is defined by the mapping of the weights from set $W$ as shown in Section~\ref{sec:graphDef}. While the significance of the method and resulting graph is the straightforward application to a variety of human factors, a few weight categories previously discussed in the literature will be defined here. Especially at the urban scale, the inclusion of varying ground conditions to walkability analysis can better predict the human experience, with comfort, timing, and fatigue being realized. 

Two cases exist for defining the cost of an edge: 
\begin{enumerate}
    \item Cost of mobility \textit{along the edge}
    \item Cost of moving \textit{to a node} 
\end{enumerate}

In the first group, metrics related to mobility itself are included; $\wm{c}$ = Cross-slope, $\wm{d}$ = Metric Distance, $\wm{s}$ = Slope, $\wm{e}$ = Energy Expenditure, $\wm{t}$ = Step Type. 

In the second group, attributes not able to be calculated by the path in isolation, but rather as values calculated cohesively with the built environment are considered. These highly localized parameters are easily missed in maximal edge connected graphs (e.g.,~\cite{lee2010computing}) and rely on the dense set of nodes throughout the environment. In this approach, an edge weight is defined by some properties or attributes of the end points (nodes) of the edge. For example, a directed edge with a start node located on concrete and end node located on sand is given a higher cost than the reverse as that node should be avoided. This weight mapping can include $\wm{k}$ = Type of Surface the end node is on, with further types; $\wm{l}$ = Light Intensity from Opposite Direction, $\wm{h}$ = Light Difference Between Nodes, $\wm{f}$ = Probability of Falling, $\wm{v}$ = Percentage of View to Greenspace, $\wm{w}$ = Surrounding Width. 

Some edge cost definitions such as the metric distance, $\wm{d}= | n_i - n_j |$, are straight forward while others $\wm{l}$ are based on the implemented simulation technique. The two most notable costs related to the focus of this paper are $\wm{c}$ and $\wm{e}$ and are therefore defined in full. $\wm{c}$ is the mapping of weights to edge costs in the graph relating to the cross slope and $\wm{t}$ is the mapping consisting of an identifier for the step type found in Alg.~\ref{alg:getConnection}. 

Calculating the cross-slope score $\wm{c}$ is done by considering the edges to $u_{ij}$ (using the mapping defined in Sec.\ref{sec:graphDef}) that are orthogonal in the $xy$ plane, denoted by ${}^{xy}\bot $. The maximum difference in height between the nodes of orthogonal edges is used as the score:

\begin{rmk}
$\wm{c} = $\\
$\texttt{max}(| \hat{z} \odot (n_{j} - n_{k}) |~\forall ~u_{ij} {}^{xy}\bot u_{ik}) ~|~ e_{ik} \in \outedges \land \wm[e_{ik}]{t} = t_{\texttt{DIRECT}}$ \\
\end{rmk}

The requirement for $\wm[e_{ik}]{t} = t_{\texttt{DIRECT}}$ corresponds to the connection type defined in Alg.~\ref{alg:getConnection} line~\ref{algln:direct} as a direct connection with no step.

Additionally, a function defining energy expenditure\footnote{This is one example, although other methods can be easily included or used in place.} of walking on slopes~\cite{minetti2002energy} as shown in Fig.~\ref{fig:energyGradient} is used for $w_e$. By correctly weighing the energy of moving both up and down a slope, human based smooth paths, rather than mathematically defined ones(e.g.,~\cite{RolesSmoothPath, Liu11Shortest}), can be found. 

\begin{rmk}\label{def:energyCost}
$\wm{e} = $\\
$280.5\wm{s}^5 - 58.7\wm{s}^4 - 76.8\wm{s}^3 + 51.9\wm{s}^2 + 19.6\wm{s} + 2.5$
\end{rmk}

Edge costs are reasonable, but often distracting or difficult to visualize in an environment. It is therefore beneficial to attribute values to a node that represent the costs associated with it. Likewise, a node may be an input mechanism for different types of simulations. In this case, the node value can be a combination of all directed edge costs away from the node (i.e., W(\outedges)), as a localized decision making agent (e.g., chemtrails used in~\cite{narahara2010self}) can be directed away from nodes that would have been difficult to leave from. The value of this node can be represented as:

$$\sum_{\pmb{\iota} \in W(\outedges)}^{}  \pmb{\iota} \odot \pmb{\rho}_{\iota}$$

Where each weight has corresponding (user defined) coefficient vector $\pmb{\rho}$ of importance.

Another case of defining an edge cost is when the existing attributes or values are first associated with a node, rather than an edge,d efined for when an edge $e_{ij}$ is assigned a weight equal to either its initial vertex ($v_i$) or terminal vertex ($v_j$). This often occurs when a value from simulation or evaluation, such as lighting, is mapped to locations in the environment, rather than the cost of moving between locations. 

\subsubsection{Path Costs}\label{sec:pathCosts}
The focus of this paper remains in the graph generation. In this section, defining various combinations and evaluations of a path is briefly discussed but is not exhaustive due to the numerous possible combinations. While a path is often defined as a sequence of vertices, a modified definition is used here to provide a simple connection to the weights described in Section~\ref{sec:edgeCosts}. As a note, the vertices of the graph are associated with nodes in the environment and therefore, the methods of path evaluation seen in space syntax such as angle or number of turns is straightforward to use. However, this section focuses on path analysis based on edge weights. 

A directed path $\Omega$ is an ordered sequence of edges that are connected in $D$ between two vertices. 

\begin{rmk}
$ \Omega := (e_1, \dots, e_m ~|~ e_n \in E(D) )$ \\
\end{rmk}

Defining the cost of an edge directly impacts the paths found when applying any common shortest path algorithm, which is synonymous with least-cost in this context. If considering the combination of distance and energy in a weighted function, this path in an urban environment can be used as the starting point for defining ramps or walkways to be constructed as to minimize the amount of land mass that must be removed. The breadth-first approach provides an efficient method in calculating this aspect of spatial evaluation as the graph generation ends when one of the accessible parameters is not met. In the indoor environment, these paths of least resistance can give direct feedback on the performance of a layout or distance between rooms. When considering the value in quantifying occupant paths of a space, multiple approaches of cost aggregation can be useful, such as distances with types of spatial views.  

By combining the weighting parameters discussed in Section~\ref{sec:edgeCosts}, the shortest path between two buildings or rooms can be quantified by more than simply able-bodied walking distances. As design alternatives and modifications are made to the environment, these values update accordingly and enable a quantified score of evaluation across the designs. When separating the costs, multiple paths can be generated for each one (or any combination there of). A space could therefore be evaluated based on the variance between each of the paths.

Additionally, edge costs can be used as the evaluation metric where, for example, the metric distance is used as the edge cost to generate least-cost paths between desired locations. Next, the weight of each edge, or the combination of the weight vector values, is applied as the evaluative score of a path, such that $S : \Omega \to \mathbb{R}$. 

\begin{rmk} \label{def:pathScore}
$$ S(\Omega) := \sum_{\pmb{w}_{\iota} \in W(\Omega)} \pmb{w}_{\iota} \odot \pmb{\rho}$$ 
\end{rmk}

In some cases the weight vector must be first processed, or alternatively calculated, to have a meaningful result. In the case of a step, $\pmb{w}$ contains an enumerator associated with connection types (Alg.\ref{alg:checkChild} line~\ref{algln:callGetConnect}). Therefore to count the total number of steps required in a path, the index of connection type $t$ in the weight vector is redefined as:

\begin{equation*}
\pmb{w}[t] = 
    \begin{cases}
    1 & \text{if } \pmb{w}[t] = t_{\texttt{OVER}} \lor t_{\texttt{UP}} \lor t_{\texttt{DOWN}}\\
    0 & \text{otherwise}
    \end{cases}
\end{equation*}

To count the number of steps in a path, $\pmb{\rho}$ would be a vectors of $0$'s, with the exception of the connection type coefficient being $1$.

\section{Implementation}~\label{sec:implementation}

The algorithms described were implemented in Python and visualized in the Rhino 3D software as it is commonly used in the free-form design and planning stages where large changes to a design based on overall accessibility can be easily made. While the same methods can be implemented in BIM focused software (e.g., Autodesk Revit), the demonstration within a modeling package containing limited information on the geometry demonstrates the scaleable use.

\begin{figure*}
\centering
    \begin{subfigure}[b]{0.23\textwidth}
    	\includegraphics[page=16, width=\textwidth]{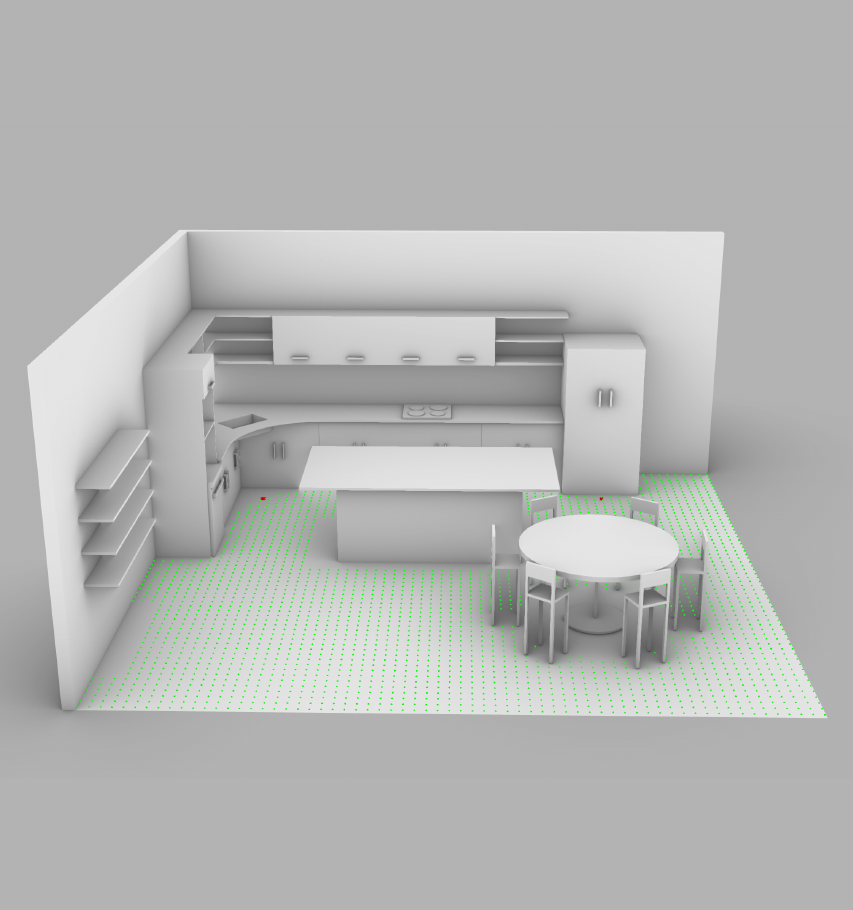}
    	\caption{}
    	\label{fig:kitchenNodes}
	\end{subfigure}
    \begin{subfigure}[b]{0.23\textwidth}
    	\includegraphics[page=12, width=\textwidth]{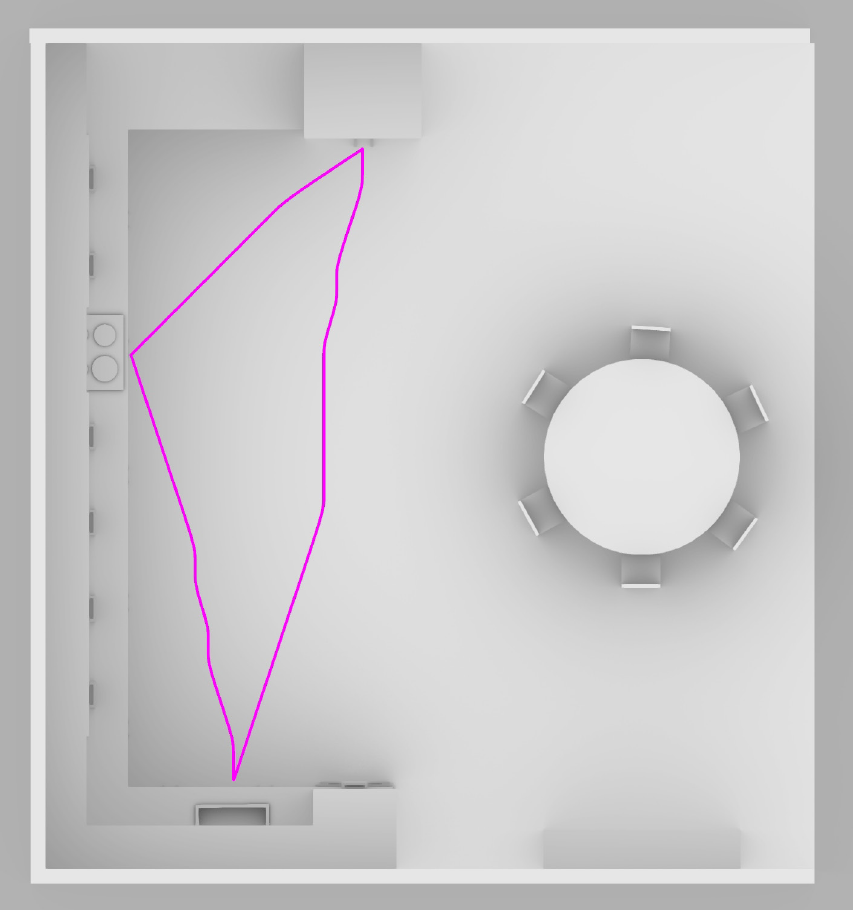}
    	\caption{}
    	\label{fig:kitchenOpen}
	\end{subfigure}
   \begin{subfigure}[b]{0.23\textwidth}
	\includegraphics[page=14, width=\textwidth]{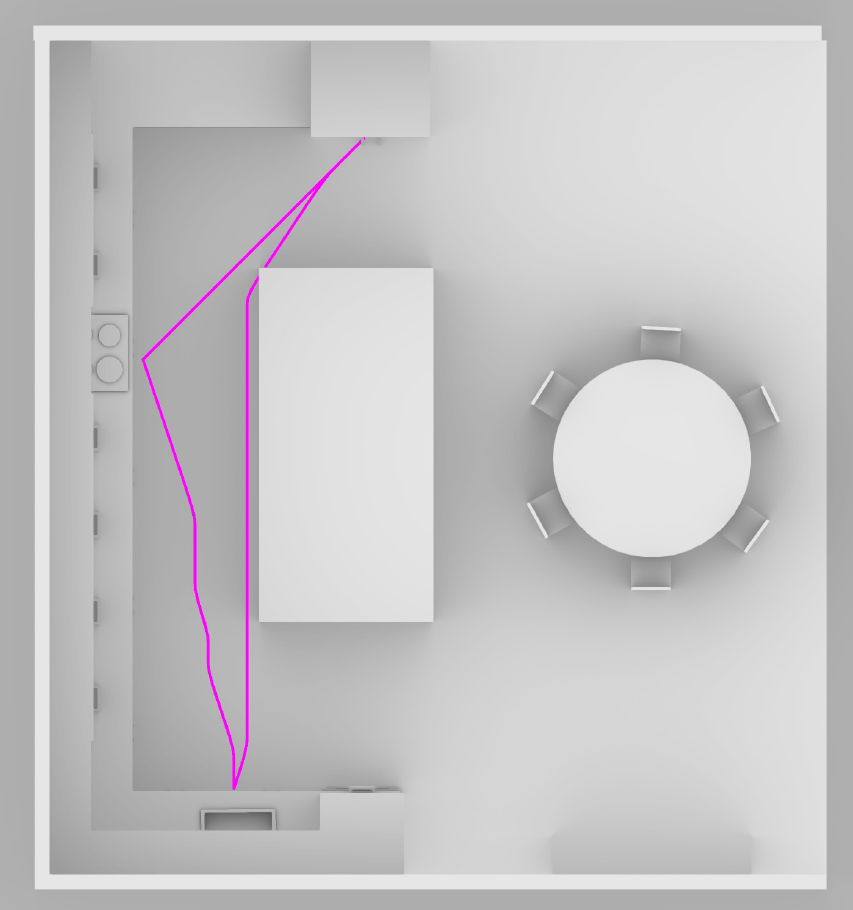}
	\caption{}
	\label{fig:kitchenIsland}
	\end{subfigure}
    \begin{subfigure}[b]{0.23\textwidth}
    	\includegraphics[page=13, width=\textwidth]{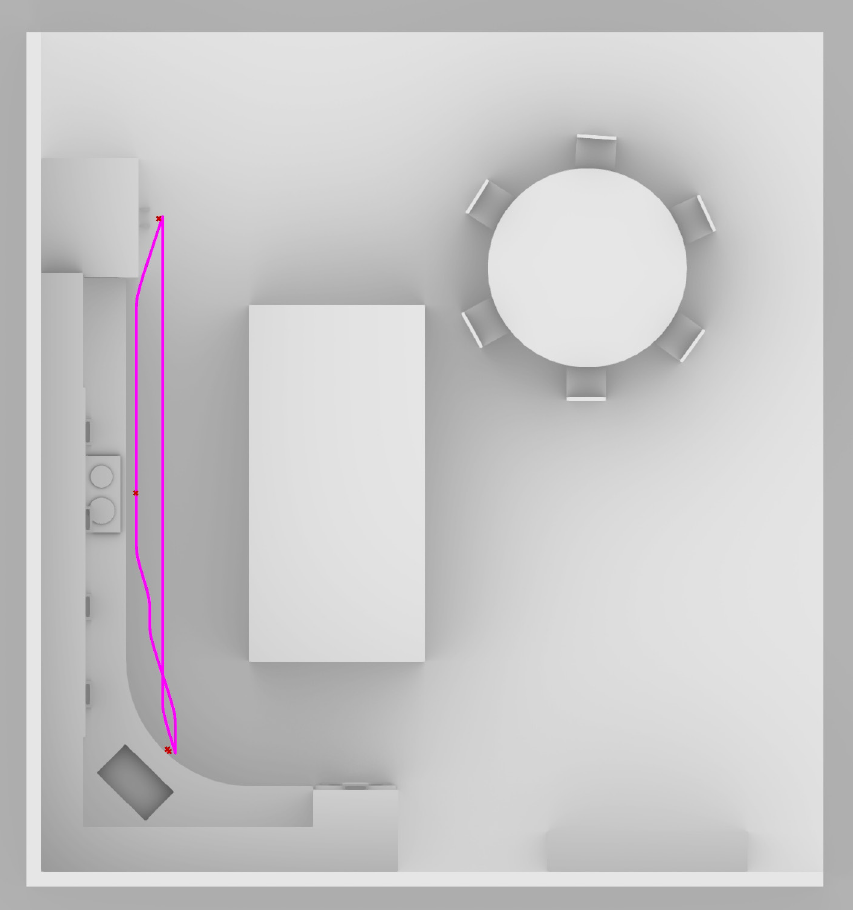}
    	\caption{}
    	\label{fig:kitchenLine}
	\end{subfigure}
    \caption{In~\ref{fig:kitchenNodes} the nodes defined as accessible locations are visualized in green. Three kitchen layouts (\ref{fig:kitchenOpen},\ref{fig:kitchenIsland},\ref{fig:kitchenLine}) are all analyzed with identical graph and path algorithms and no modification to the parameters with a path between the sink, stove, and refrigerator. ~\ref{fig:kitchenOpen} shows a typical work triangle of a kitchen. In~\ref{fig:kitchenIsland}, the path between the refrigerator and sink is interrupted by an island, diverting the path around it. ~\ref{fig:kitchenLine} shows a smaller and modified layout where the path between the key points is nearly a straight line. } 
    \label{fig:kitchen} 
\end{figure*}

The code was developed in Python and IronPython, with a SQL database to manage graph data between them. During the graph creation a Python Dictionary object is used to store the vertex and edge connections, which is later converted to a compressed sparse row (CSR) structure with multiple arrays set as data for costs. To demonstrate the algorithms in various cases a simple GUI was made with the Rhino Grasshopper interface. No modifications to the algorithm are made between examples, except for any human factor input parameters that are used for demonstration purposes. The combination of these examples illustrates the flexibility in design use-cases and robustness at varying levels of geometric detail.

While more concrete examples of the algorithm are given in the following sections, Figure~\ref{fig:Overview_multiLevelPath} provides an overview of the process from a design-user perspective. Given an initial building model, the user would select a location to start the algorithm from. Selecting the metrics of interest, such as the open space around each location, a heatmap visualizing the data can be applied to the nodes of the graph. Either the nodes themselves can be explored, or the nodes with the building model can be seen in the modeling program at the same time (or nodes can be hidden all-together). Each image in the diagram is taken as a screen capture from the viewport, with only the text box and start/end waypoints overlaid for illustration purposes. 

To demonstrate the feasibility, a \ac{AG} was generated on three test cases: a kitchen representing the object scale, the NJIT Architecture building for the building scale, and two height varying surfaces representing terrain types. The intent of these examples is to clearly and understandably communicate the flexibility, performance capability of the algorithm presented in this paper, and the direct impact quantitative metrics have on the generated graph. Both the kitchen (Sec.~\ref{subsec:kitchen}) and building (Sec.~\ref{subsec:building}) models are exported from an Architects Revit and Interior Designers Rhino 3D models (Adapted from~\cite{Schwartz:2018:HTD:3289750.3289768}). In Section~\ref{subsec:topo}, both a real-world GIS terrain is shown and a more simplified model is used to help readers in understanding how varying terrain types impact energy costs and can be included in path planning and spatial analysis. 

\subsection{Demonstration on Room Scale}\label{subsec:kitchen}

At the room scale, a kitchen provides a demonstration of the graph avoidance of objects. A common design process for a kitchen is the evaluation of a \textit{work triangle} defined by the refrigerator, sink, and stove (store, preparation, and cooking). By selecting these objects as destinations for a path, this triangle can be generated and evaluated automatically. The object level inclusion of the graph enables kitchen islands to be included in the evaluation rather than considering the direct path between the objects of interest (e.g., stove).

Using the accessibility graph, Figure~\ref{fig:kitchen} shows three basic kitchen layouts. The table and chairs, along with counters are not included in the accessible locations. The stove, refrigerator, and sink positions are selected to calculate the travel between them, creating the \textit{work triangle}. Using the distance metric of the path, the variation in the triangle perimeters is simply calculated. As the path is created from the nodes, an exact shortest path can vary based on the distance a node is connected to its neighbors. In Figure~\ref{fig:kitchenOpen} the total calculated distance is 1091 cm, while the exact distance is 1080 cm. The difference between the estimated and actual distances is 11cm, or 1\%. Individual paths are calculated as follows (calculated vs actual): Stove-Refrigerator 242 cm vs 241 cm, Stove-Sink 343 cm vs 340 cm, Sink-Refrigerator 506 cm vs 500 cm. 

In comparison, to Figure~\ref{fig:kitchenOpen}, an island blocking the direct (Fig.~\ref{fig:kitchenIsland}) path changes the total distance from 1091 cm to 1114 cm. With the island blocking the path, the euclidean distance between the sink and refrigerator is calculated as 532 cm compared to the original 506 cm. Finally, the smaller kitchen arrangement of Figure~\ref{fig:kitchenLine} provides a total perimeter of 830cm for the \textit{work triangle}. While these values represent an isolated test case for demonstrating the method in implementation, the example is in the context of a commonly defined approach to interior design. In the simplest form, this evaluation provides instant feedback as to the distances between objects in the \textit{work triangle} in which designers may want to minimize for a desired occupant experience. In addition to traveling distances, a heatmaps could be visualized for the designer to highlight areas that are dangerous for an occupant to fall, or locations difficult to reach certain cabinets, as illustrated in~\cite{Schwartz:2018:HTD:3289750.3289768}.  

\begin{figure*}
    \centering
    \begin{minipage}{.32\linewidth}
            \begin{subfigure}[t]{.99\linewidth}
                \includegraphics[page=17, width=\textwidth]{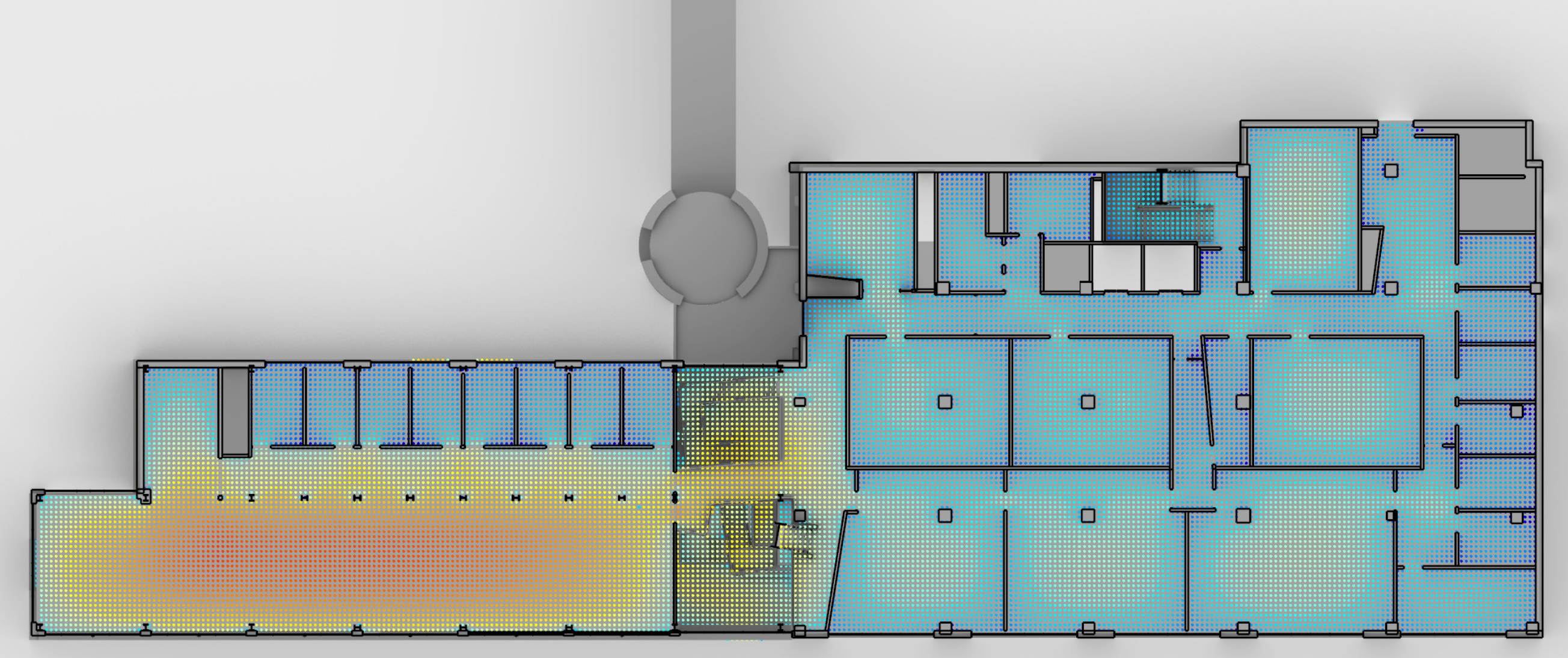}
                \caption{Heatmap of Open Space}
                \label{fig:buildingPath_Heatmap}
            \end{subfigure}
            \begin{subfigure}[t]{.99\linewidth}
                \includegraphics[page=18, width=\textwidth]{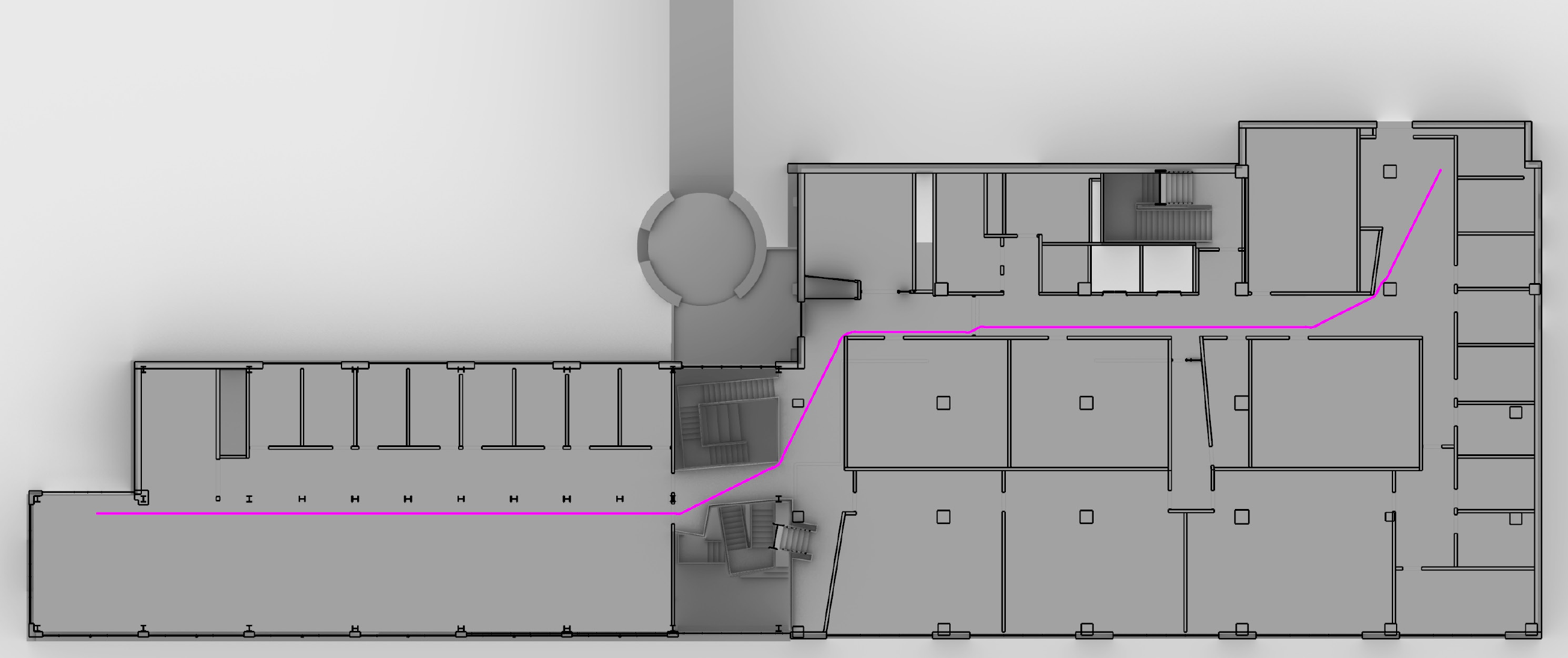}
                \caption{Minimize Distance}
                \label{fig:buildingPath_dist}
            \end{subfigure}
        \end{minipage}
    \begin{minipage}{.33\linewidth}
        \begin{subfigure}[t]{.99\linewidth}
            \includegraphics[page=20, width=\textwidth]{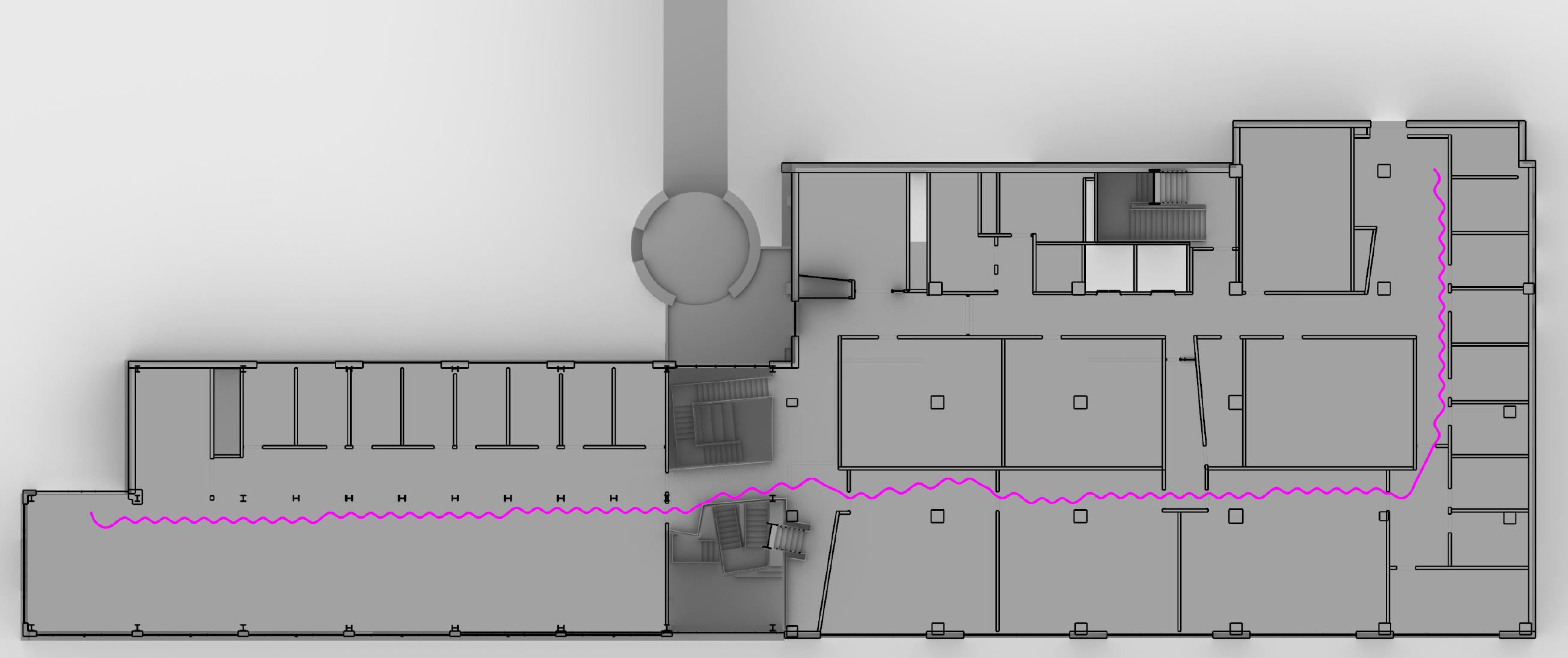}
            \caption{Maximize Open View}
            \label{fig:buildingPath_MaxOpen}
        \end{subfigure} \\
        \begin{subfigure}[t]{.99\linewidth}
            \includegraphics[page=24, width=\textwidth]{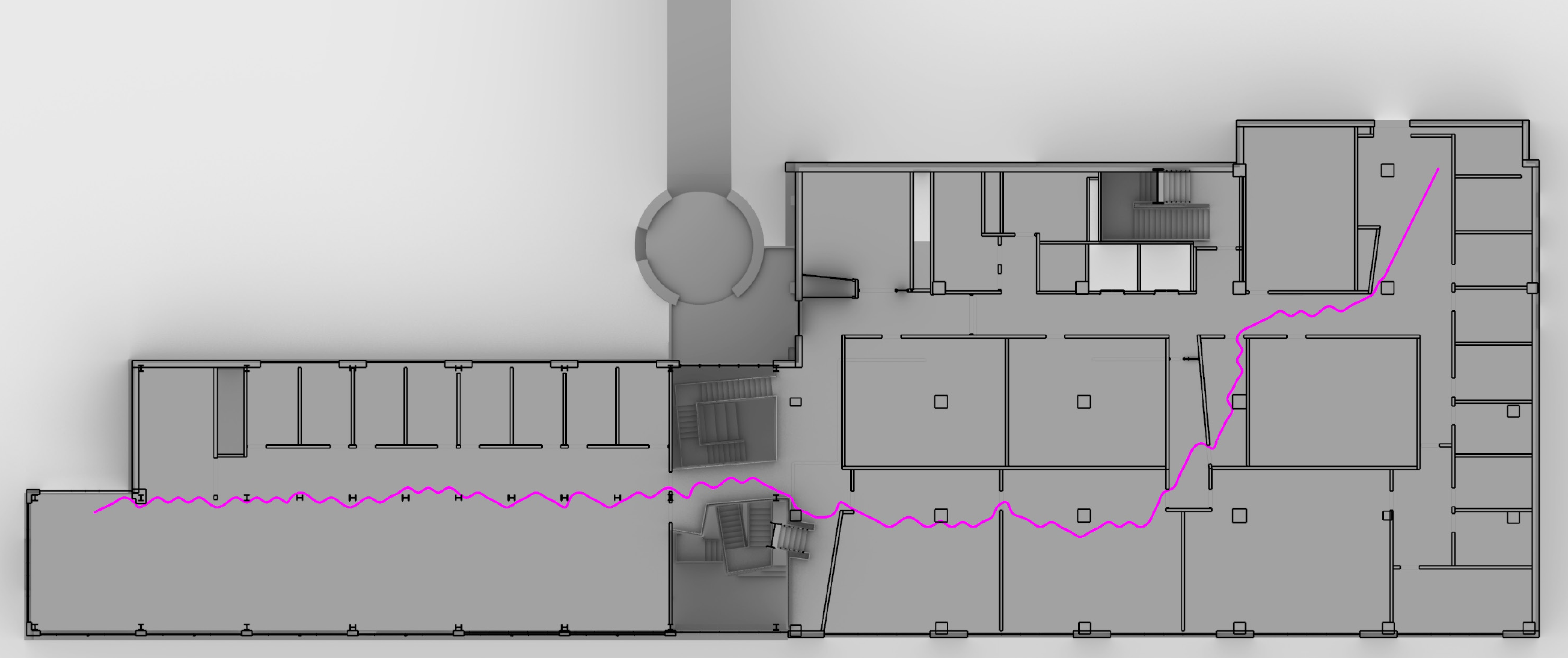}
            \caption{Avoid Open View}
            \label{fig:buildingPath_AvoidOpen}
        \end{subfigure} \\
    \end{minipage}
    \begin{minipage}{.33\linewidth}
        \begin{subfigure}[b]{.99\linewidth}
            \includegraphics[page=21, width=\textwidth]{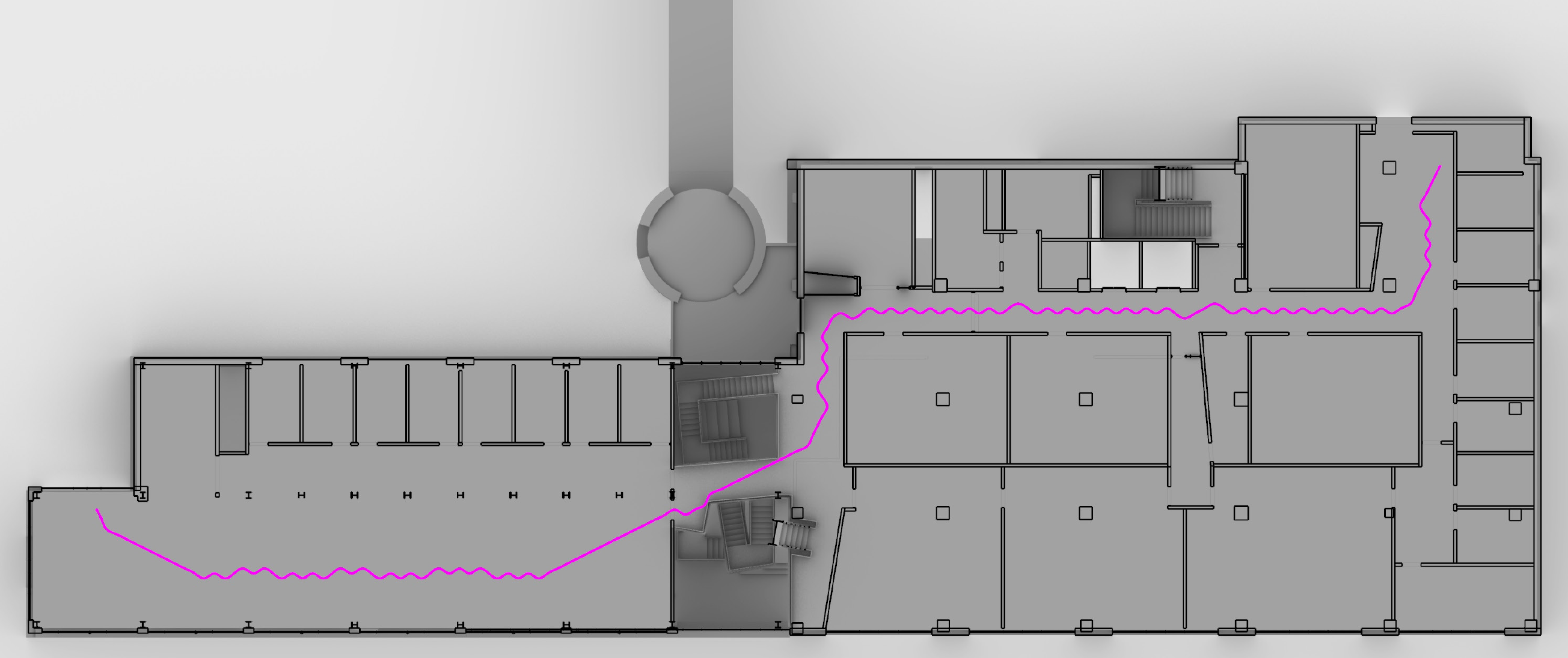}
            \caption{Avoid Narrow View}
            \label{fig:buildingPath_AvoidMin}
        \end{subfigure} \\
        \begin{subfigure}[b]{.99\linewidth}
            \includegraphics[page=23, width=\textwidth]{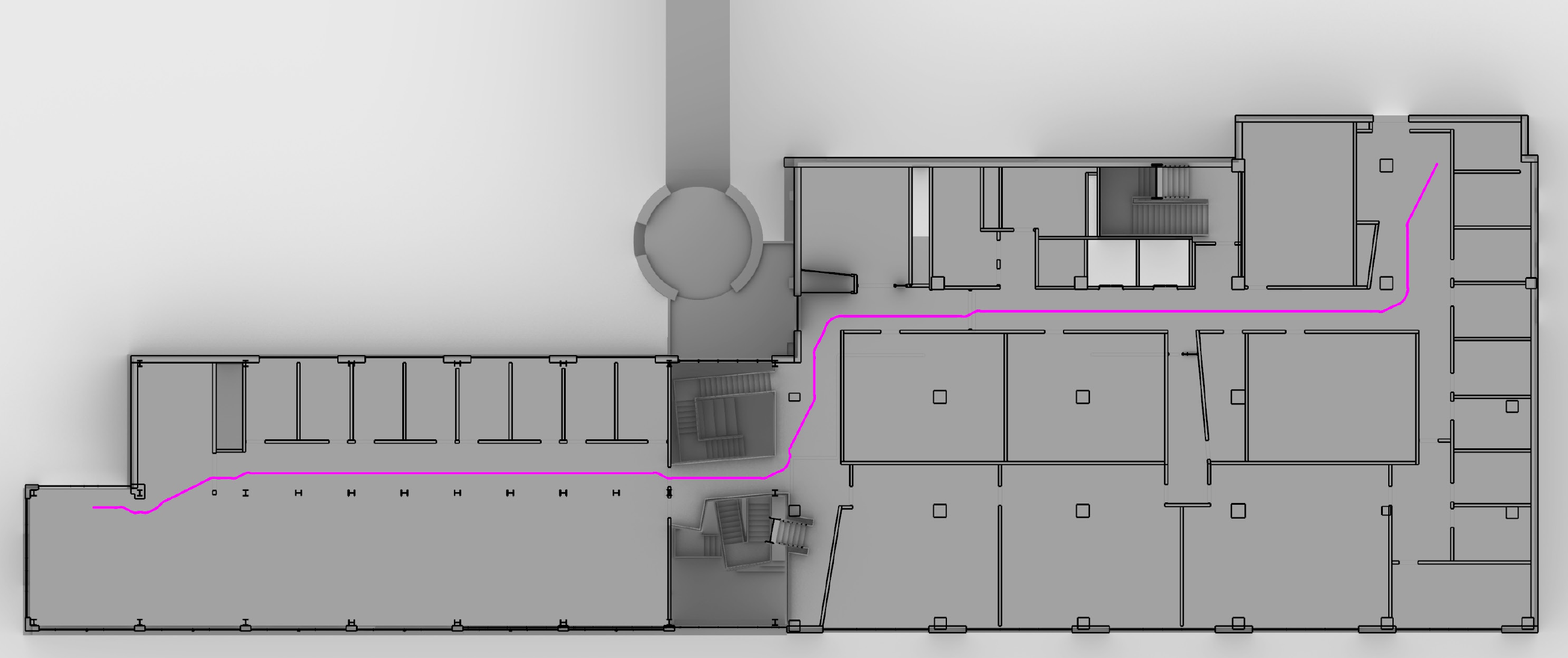}
            \caption{Minimize Distance Avoiding Obstacles}
            \label{fig:buildingPath_DistView}
        \end{subfigure} 
    \end{minipage}
    \caption{Demonstration of different paths created with the same graph using different and combined attributes. \ref{fig:buildingPath_Heatmap} shows a heatmap visualization of the average distances in a viewshed analysis, where red is the most open views and blue the least open views. \ref{fig:buildingPath_dist} is a standard shortest path using metric distance. \ref{fig:buildingPath_AvoidOpen} avoids paths based on the maximum view while~\ref{fig:buildingPath_MaxOpen} is the inverse. In comparison,~\ref{fig:buildingPath_AvoidMin} avoids edges based on the minimum view distance. Finally,~\ref{fig:buildingPath_DistView} combines the edge costs of open space (\ref{fig:buildingPath_AvoidMin}) and distance (\ref{fig:buildingPath_dist}) by setting a multiplier to edges within a threshold close to obstacles, creating a wall-buffer effect.}
    \label{fig:pathView}
\end{figure*}

\subsection{Demonstration on Building Scale}\label{subsec:building}

When descritizing the model to nodes of accessible locations, certain features such as proximity to a wall are not accounted for when using a shortest path algorithm. In fact, this is one motivation for the importance of numerous quantitative human factors being included in the graph as the majority of existing work either: removes graph nodes near obstacles to force paths away (i.e., the walls), uses an extremely sparse graph containing predefined locations (e.g., center of doorways), or ignores the issue by interpreting the data as an approximation. When removing nodes, the graph loses data that exists for human occupants to solve only the pathing similarity problem. The importance of using data to determine a path or evaluating a shortest path was elaborated on in Section~\ref{sec:humanFactors}. By predetermining how occupants will move or experience an environment, the evaluation of that movement loses meaning. While these options are the best (only) ones available, the ongoing efforts of human-building interaction and cognitive science enables us to more accurately model and simulate environmental conditions of an occupant. 

By no means is spatial visibility and distance the only two conditions considered by an occupant. However, Figure~\ref{fig:pathView} demonstrates the diversity of paths between identical locations when only two quantitative metrics are used in the \ac{AG}. In this demonstration, a viewshed analysis is performed at a height of 1.8 meters using 2000 rays per location and a 360 degree field of view, with $\pm40^{\circ}$ elevations. The \ac{AG} was generated at a grid spacing $a$ of 25cm with a minimum connectivity of 8 nodes, which the result can be seen in the top right empty rooms (Fig.~\ref{fig:buildingPath_Heatmap}) that have an opening too small to be considered accessible. For each location, the maximum and minimal distance any ray travels in the set is recorded and used as node attributes. These values are then set as the outgoing edge cost (incoming costs would provide similar results) to that node. The use of maximal or minimal views are not suggested as realistic paths, but rather as a demonstration of how these metrics can find a variety of pathing strategies. Additionally, any metric can be calculated along any path. 

In Figure~\ref{fig:buildingPath_dist} the shortest path by distance between two locations is visualized, totaling ~78.7 meters. Another example is to use the farthest distance visible from a location as the edge cost, which would create high costs for any edge that has a long line of sight. This approach creates the path in Figure~\ref{fig:buildingPath_AvoidOpen}, which is ~90.2 meters long. The reciprocal score ($\frac{1}{\texttt{max\_dist}}$), where edges cost less when there is a distant object (i.e., longer line of sight) is shown in Figure~\ref{fig:buildingPath_MaxOpen} at length of ~95.5 meters. Rather than avoiding or using edges based on the longest distance, the same can be applied for the shortest (i.e., closest distance to an object in view). Similarly to the previous method, close distances can be avoided by dividing 1 by the minimal distance, shown in Figure~\ref{fig:buildingPath_AvoidMin}, with a path length of ~90.9 meters. A final path is shown in Figure~\ref{fig:buildingPath_DistView} that combines the edge costs of distance and a function of minimal views($\Omega$ in Sec.~\ref{sec:pathCosts}). By setting a threshold of 63cm, any outgoing edge from the node is set to the length of the edge multiplied by a constant (4, in this example). This is, practically, making an edge that moves within 63cm of a wall cost four times as much as traversing an edge further away.  By doing this, a minimal distance path is created that avoids obstacles (e.g., walls, handrails) without needing to remove nodes from the graph of accessible locations. This combined-cost path totals ~82.2 meters, ~3.5 meters longer than the minimal distance and ~8.7 meters shorter than the path purely avoiding obstacles. 

\begin{figure*}
  \centering
   \includegraphics[page=16, width=\textwidth]{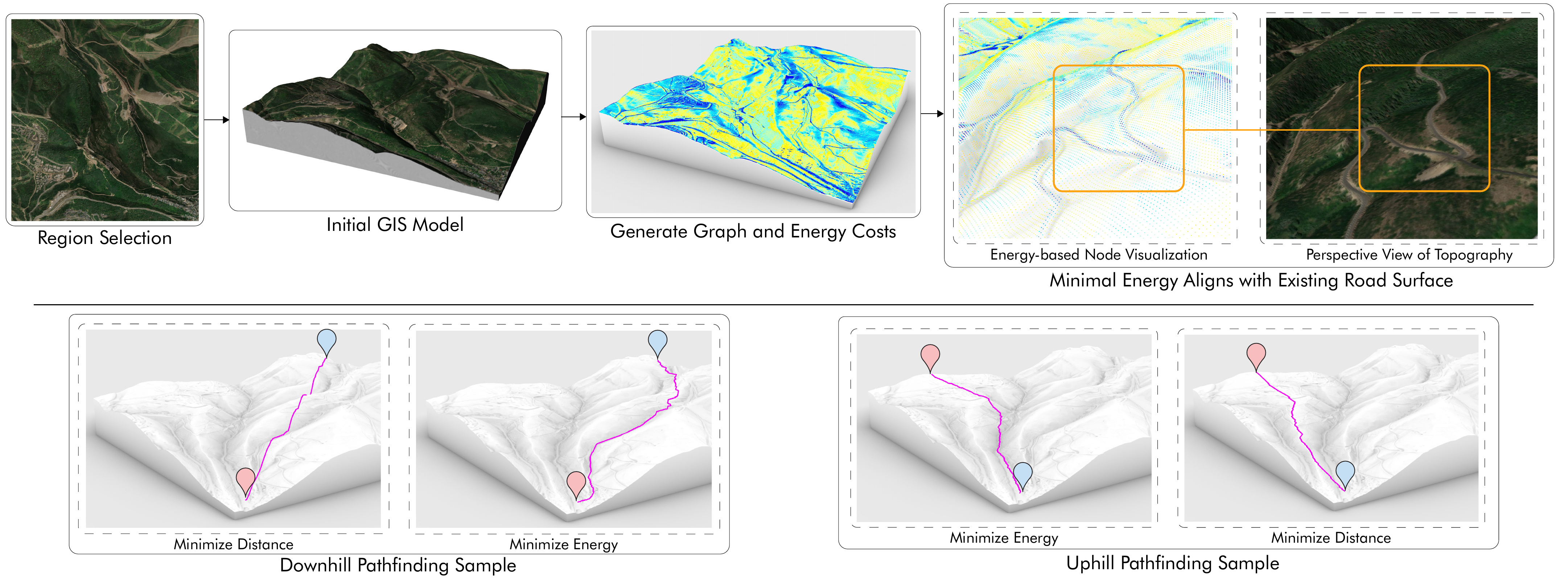}
  \caption{(Top) Demonstration of using GIS data as input model for generating graph and analysis. Nodes are visualized as a heatmap from low to high energy cost (blue to red). (Bottom) Paths are demonstrated from start-point (blue pin) to end-point (red pin) based on energy and distance.} 
  \label{fig:gisPath} 
\end{figure*}

\subsection{Demonstration on Topography}\label{subsec:topo}

At the largest scale, Figure~\ref{fig:gisPath} shows a process in which GIS data was used to visualize areas of high and low energy expenditure, as well as paths generated between locations that minimize either distance or energy. In this example, a region near Park City, Utah was selected within CityEngine~\cite{cityEngine} at a size of approximately $2,500m^2$. In this example, the limitation of the algorithm is not the size, but rather the model resolution. The accuracy at which the topography can be created limits the meaning of certain conditions, such as slope and step. For example, while there are clearly trees in the mountains, the 3D model itself lacks these geometries. The geometry can still be analyzed, but two conditions are applied: (1) The margin of error for specific metrics such as energy expenditure is assumed to be large (2) Spacing of the nodes is sufficiently large so that errors from tessellated ground conditions are minimized. In this example, node spacing was set to 5 meters, across which a maximum slope of $\pm 45^{\circ}$ was allowed. Looking at the node visualization it becomes apparent there existing pathways that have been evaluated as \textit{low energy}. Upon closer inspection, while the the node spacing is larger than a human step, roadways that have been carved into the mountain are still captured at the course resolution and are found by the \ac{AG}. As the complexities of the large-scale terrain are difficult to intuitively understand, two simple examples are additionally provided to clearly demonstrate the role of slope and cross-slope. 

\begin{figure}[pos=h]
\centering
    \begin{subfigure}[b]{0.23\textwidth}
    	\includegraphics[page=3, width=\textwidth]{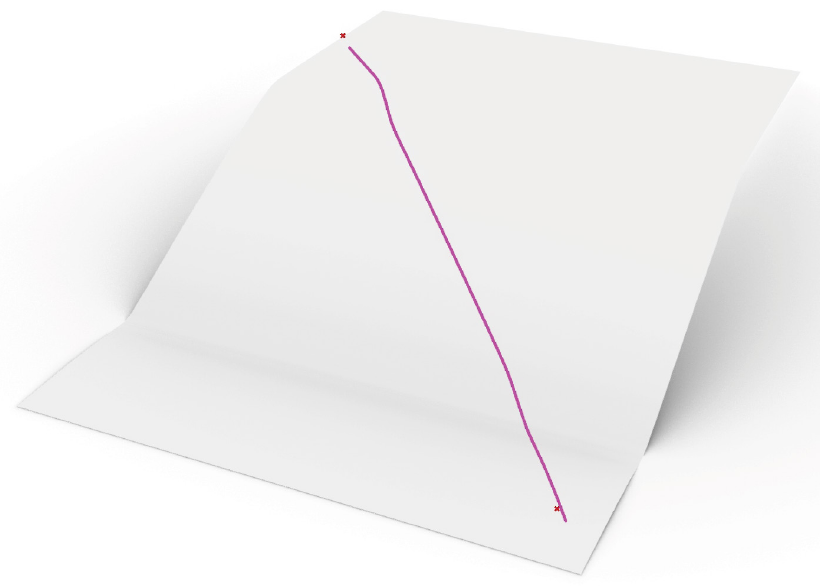}
    	\caption{Distance}
    	\label{fig:distPath}
	\end{subfigure}
    \begin{subfigure}[b]{0.23\textwidth}
    	\includegraphics[page=4, width=\textwidth]{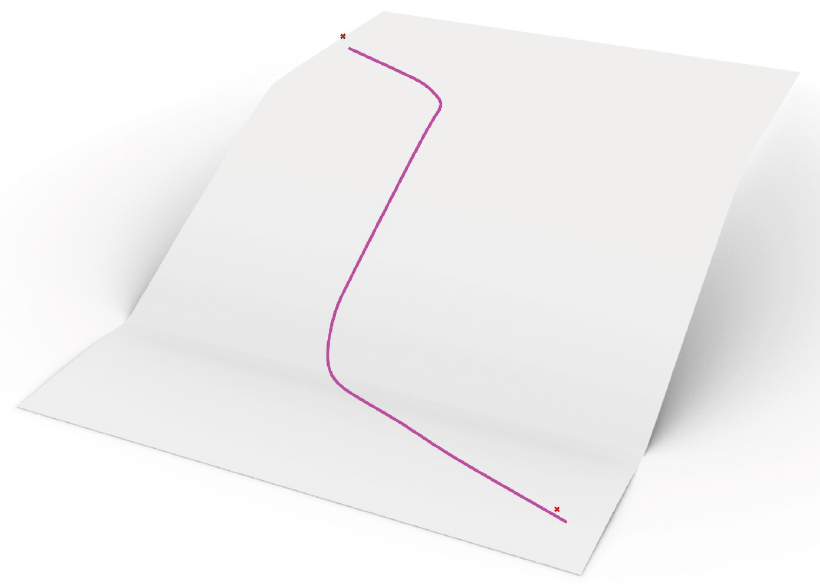}
    	\caption{Cross-slope}
    	\label{fig:crossPath}
	\end{subfigure}
    \caption{Two paths generated with (\ref{fig:distPath}) distance only and (\ref{fig:crossPath}) distance with the added cost of a cross-slope. } 
    \label{fig:pathCross} 
\end{figure}

Cross-slope avoidance in path generation is demonstrated in Figure~\ref{fig:pathCross}. In Figure~\ref{fig:distPath}, two corners of a ramp are used as start and end points of the path, with a shortest path metric of distance being used. When the angle of the cross-slope is added as an additional weight to the distance, the shortest path generated follows the ramps incline and moves diagonally on the flat surface (Fig.~\ref{fig:crossPath}). 

\begin{figure}[pos=h]
\centering
    \begin{subfigure}[b]{0.23\textwidth}
    	\includegraphics[page=6, width=\textwidth]{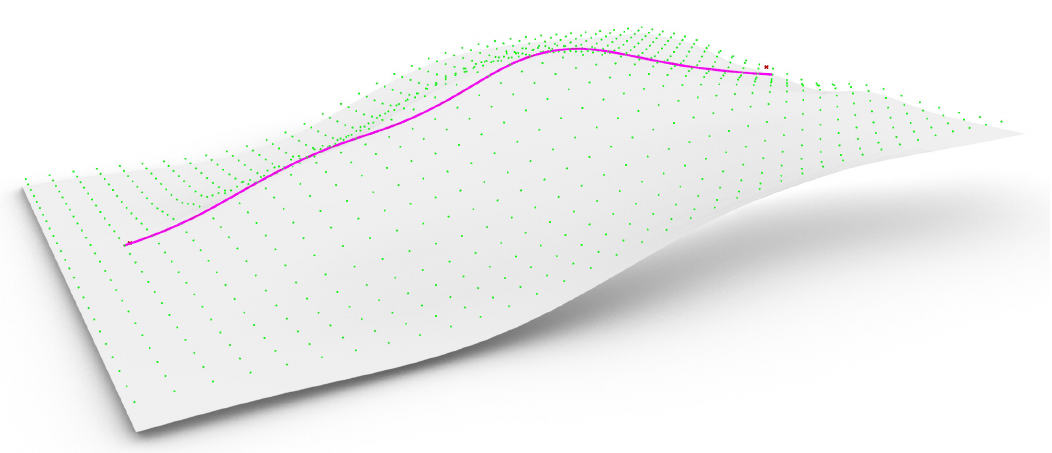}
    	\caption{Distance}
    	\label{fig:pathEnergyDist}
	\end{subfigure}
    \begin{subfigure}[b]{0.23\textwidth}
    	\includegraphics[page=5, width=\textwidth]{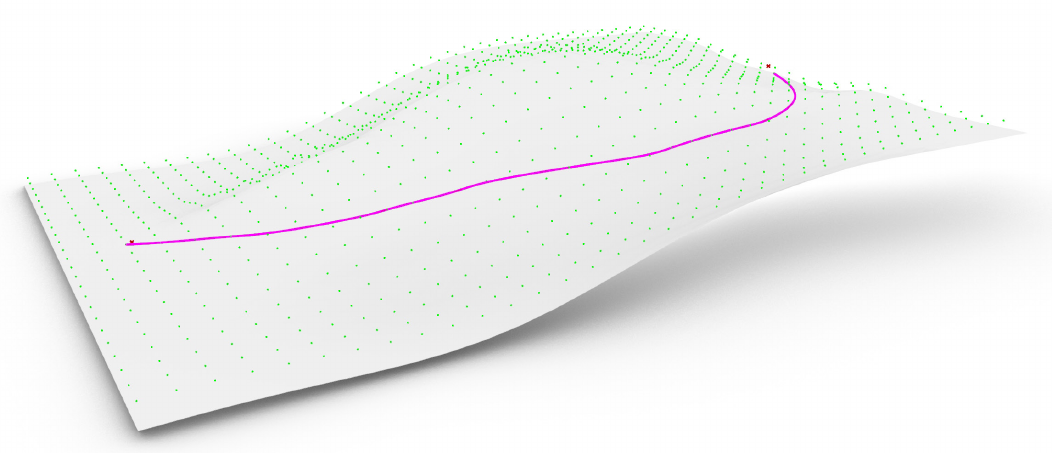}
    	\caption{Energy}
    	\label{fig:pathEnergyEnergy}
	\end{subfigure}
    \caption{Two paths generated with (\ref{fig:pathEnergyDist}) distance only and (\ref{fig:pathEnergyEnergy}) energy. } 
    \label{fig:pathEnergy} 
\end{figure}

As the edge costs of various factors can be either the method for finding the shortest path, or for generating it, there are two evaluative scores. In~\ref{fig:pathEnergyDist}, the shortest path is created using the edge costs of distance in which the value is 64m and the energy cost is 470 J/KG m. In~\ref{fig:pathEnergyEnergy}, energy is used as the edge cost, reducing the energy calculation to 355 J/KG m, with a distance of 73m. This exemplifies the complex relationship in landscape environments for human mobility as a 14\% increase in path distance results in a 24\% decrease in energy.

Both the visualization and data can be used in the planning of environment paths and site selection. By understanding the physical impact the topography has on people, designers can include in their planning a quantitative metric of consideration for placing entrances and exits in a way that may improve accessibility and reduce the need for large changes to the natural environment. 

\section{Discussion}~\label{sec:discussion}

The consideration of fatigue and energy in an environment is greatly applicable in urban environments and decisions as to the reduction of hills and soil, a costly construction process. The results show how a highly complex topography can be analyzed and a minimal-energy path (which has a non-linear relationship with slope), is automatically found. Indoor, this type of calculation has important implications for large spaces such as airports, and with the integration of indoor and outdoor, environments such as a school campus can use this type of automated evaluation method for improving and determining accessibility and timing between classes (i.e., large scale built environment circulation analysis). 

The visualization method of heatmaps shown in Figures~\ref{fig:buildingPath_Heatmap} and~\ref{fig:gisPath} is a commonly used technique for providing intuitive feedback for a designer. Likewise, path visualizations (Figs.~\ref{fig:kitchen}-\ref{fig:pathEnergy}) are commonly used in circulation analysis. In either technique for providing feedback to the designer, the important aspect is in what data is being visualized. By using the \ac{AG} method, not only can existing work (e.g., visibility graphs) be used with its base graph structure, but high-fidelity human-centric data relating to physiology can be at the forefront of the design. With construction techniques pushing the limits of creation in the built environment, such as creating curved concrete floors and ramps, it is important to provide tools that can quantitatively evaluate how a human will physically experience and navigate such a design.  

There are multiple ways to approach the navigable or accessible areas in a 3D model, each with their own drawbacks.  While a process using voxelization could under-include geometry (e.g., ignoring geometry below the defined voxel size), it could likewise over-include these small details.  As the method presented in this paper uses ray intersections, a point of failure in accurate representation of accessible locations could be a model containing extremely small geometry such as a metal wire across the room. Future work to resolve this could include the use of geometry bounding boxes relative to the edge connection of nodes. 

The flexibility in the proposed algorithm for use on non-labeled data is similarly limited in instances where there is no guaranteed way to infer movable structures from the 3D model, such as the case of elevators. This is an inherit issue with all methods, and extending this work to utilize the same techniques (e.g., defining a node at the landing of the elevator on each floor) would be straightforward. The \ac{AG} could be generated on each floor, with a labeled elevator (and hence, closest node) connecting the floors. Furthermore, the step height parameter could be set to 0, illustrating through path visualization an occupant using a wheelchair would have limited accessibility.  It is still important to note, however, that the proposed method excels without labeled data and is able to traverse 3D environments on varying surfaces and stepping blocks (e.g., stairs) even when these objects are not labeled as such, which would likewise highlight to a designer that subsequent floors were not accessible without elevators (i.e., missing a ramp).  

Changing the user input parameters can allow for multiple specific graphs to be analyzed, for example, useful for toddlers/children in childproofing a room by setting the height small enough for a child and see if they can access unsafe space, or detect sharp edges of objects within their path, or give a metric for them climbing on furniture. Likewise, an extension to the weighting parameters of discrete evaluation metrics defined for edge costs, an additional dimension of the weight vector could be added to incorporate time.  This could be used, for instance, in considering factors such as lighting levels where additional edge costs would be generated across times of the day. 

While this research focuses on presenting a method for generating a graph and then using the graph for path-based analysis, the human factors discussed, and underlying graph representation, have applications in related research areas. Although a navigation mesh could be more efficient if the only goal was agent-based simulation, the digraph could be used as a layer for both navigation policy and as a position-aware data structure for storing information on agent interactions (e.g., chemical diffusion rates for agent modeling~\cite{narahara2010self}). In furniture layout optimization procedures~\cite{yu2011make,weiss2017automated}, an input metric is needed to drive the weighting factors of different configurations. Although early work in this area used object relationships such as couch to television~\cite{yu2011make}, future research could leverage the graph representation embedded with human factor data at each node to include more complex dynamics, such as visual noise calculated by the intersection of multiple objects within a view frustum. 
Beyond discrete evaluation metrics, this research, and corresponding in-depth literature review, will ideally reinforce the need for considering numerous human factors in all aspects of computational design. Ideally, the increased use of machine learning and optimization methods for the built environment and design will utilize evaluative data about people (i.e., variation in floor surface and accessibility) and as such include the human more, not less, in the design of a space.  

\section{Conclusion}\label{sec:conclusion}

In this paper, a method for creating a weighted and directed graph (digraph) from a 3D model with a focus on human-relevant attributes for traversing a space, referred to as \ac{AG}, was presented. The process for generating the graph uses ray-casting on the data structure used for other purposes in CAD, such as a Bounding Volume Hierarchy (BVH), which improves performance and minimizes the time it takes for design-analysis. \ac{AG} was created within environments containing multiple levels and stairs, as well as on varying heights such as slopes, cross-slopes, and other complex surface conditions, demonstrating unique paths based on distance, energy, and numerous visibility metrics--not feasibly found manually by a designer. Not only does \ac{AG} work on varying ground conditions, but includes the information as data attributes in the graph (e.g., node connections as a step or incline) which enable high fidelity accessibility checks to include physiological effort and other human-experiences. 

In summary, the following benefits to the integration of various design processes are achieved: 1) Does not require extensive pre-processing such as voxelization, making the query-result process faster. 2) Does not require (but can incorporate) labeled data such as BIM, enabling earlier-stage feedback (e.g., with mass models). 3) Creates a Graph representation rather than general surface or polygon bounds, enabling spatially high-fidelity metrics to be incorporated. 4) Only includes human-accessible locations in the graph, reducing computation time as well as highlighting areas of inaccessible locations. 5) Works for both indoor and outdoor environments regardless of building levels. 6) Defines and calculates edge costs by human factors, including slope, cross-slope, and steps-- providing a higher resolution than past works of occupant experience and enabling more accurate accessibility analysis for differently-abled occupants. 7) Easily used with existing graph theory (e.g., visibility graphs) and design evaluation methods such as lighting as to not replace the vast existing literature, but instead enhance and augment it. 

This work can aid designers and researchers in automated analysis of the built environment for physiological metrics of varying types of occupant mobility. \ac{AG} can be a platform for others to quickly and easily generate graph representations of the space and integrate their own additional metrics, useful in facilitating quantitative evaluations of interest, and facilitating human-centric designs.

\section{Acknowledgements}

This paper contains work that was supported in part by the U. S. Army Combat Capabilities Development Command (CCDC) Armaments Center and the U. S. Army ManTech Office under Contract Delivery Order W15QKN19F0002 - Advanced Development of Asset Protection Technologies (ADAPT).




\bibliographystyle{autocon-format}

\bibliography{cas-refs}

\appendix
\section{Appendix}\label{appendix}
A video screen recording of the breadth-first generation of nodes is provided in the supplemental materials.

\begin{figure*}
\centering
    \begin{subfigure}[b]{0.32\textwidth}
    	\includegraphics[page=6, width=\textwidth]{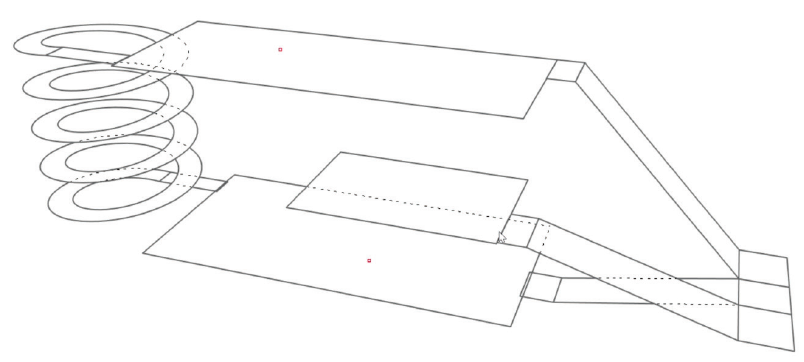}
    	\caption{}
    	\label{fig:bfs1}
	\end{subfigure}
    \begin{subfigure}[b]{0.32\textwidth}
    	\includegraphics[page=7, width=\textwidth]{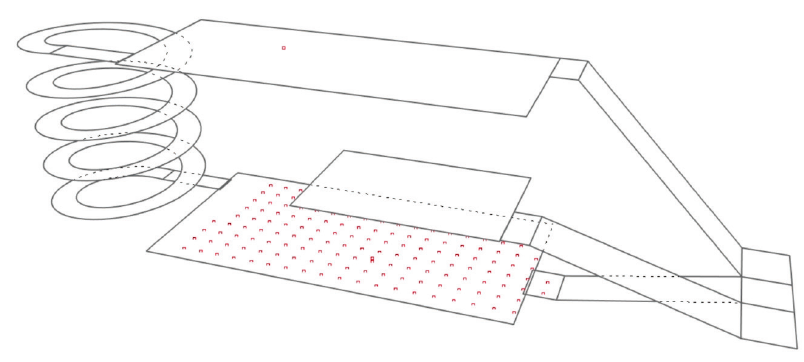}
    	\caption{}
    	\label{fig:bfs2}
	\end{subfigure}
    \begin{subfigure}[b]{0.32\textwidth}
        \includegraphics[page=8, width=\textwidth]{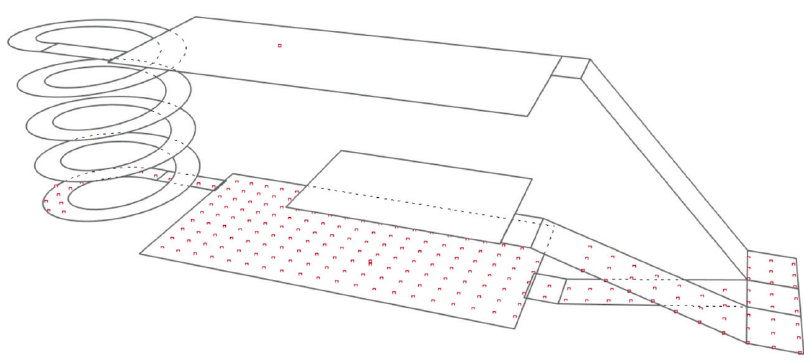}
        \caption{}
        \label{fig:bfs3}
    \end{subfigure}
    \begin{subfigure}[b]{0.32\textwidth}
    	\includegraphics[page=9, width=\textwidth]{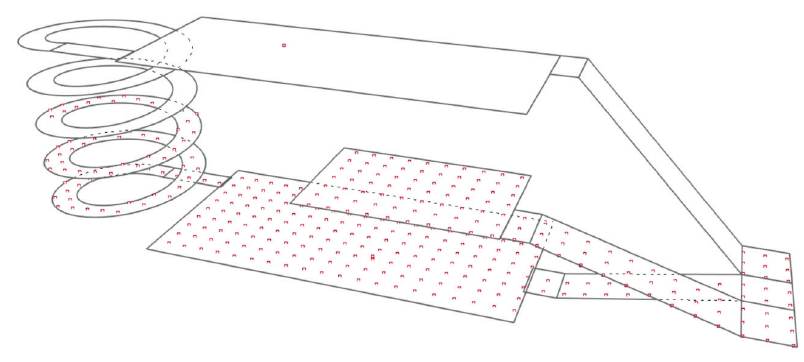}
    	\caption{}
    	\label{fig:bfs4}
	\end{subfigure}
    \begin{subfigure}[b]{0.32\textwidth}
    	\includegraphics[page=10, width=\textwidth]{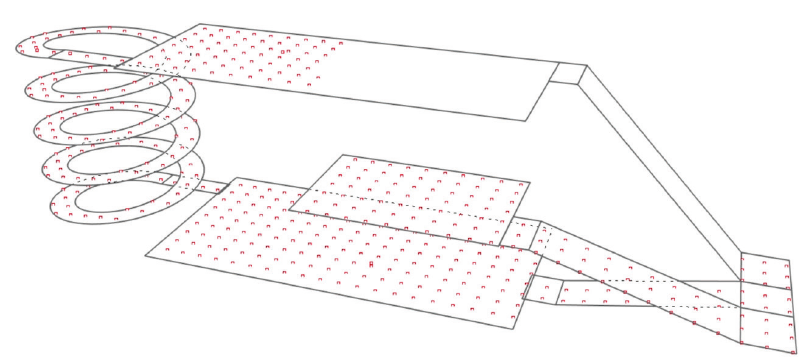}
    	\caption{}
    	\label{fig:bfs5}
	\end{subfigure}
    \begin{subfigure}[b]{0.32\textwidth}
    	\includegraphics[page=11, width=\textwidth]{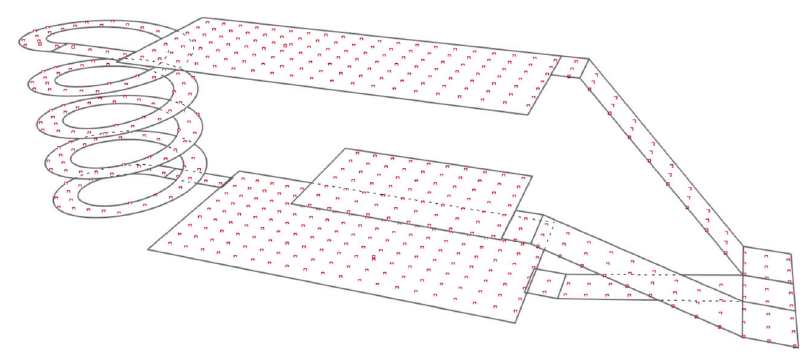}
    	\caption{}
    	\label{fig:bfs6}
	\end{subfigure}
    \caption{Example stills of the supplemental video showing the growth of the accessible node locations. In~\ref{fig:bfs3}, the breadth-first search for accessibility has stopped at the steep ramp on the right side due to the input user parameters, while subsequent frames of~\ref{fig:bfs5} and~\ref{fig:bfs6} show the graph continuing on the spiral ramp and connecting on the down ramp. As the graph is directed a shortest path from bottom to top would use the spiral ramp, a path from top to bottom would use the ramp on the right.} 
\end{figure*}



\end{document}